\documentclass[letterpaper]{article} % DO NOT CHANGE THIS
\usepackage{aaai24}  % DO NOT CHANGE THIS
\usepackage{times}  % DO NOT CHANGE THIS
\usepackage{helvet}  % DO NOT CHANGE THIS
\usepackage{courier}  % DO NOT CHANGE THIS
\usepackage[hyphens]{url}  % DO NOT CHANGE THIS
\usepackage{graphicx} % DO NOT CHANGE THIS
\usepackage{natbib}  % DO NOT CHANGE THIS AND DO NOT ADD ANY OPTIONS TO IT
\usepackage{caption} % DO NOT CHANGE THIS AND DO NOT ADD ANY OPTIONS TO IT
\usepackage{bm}
\usepackage{amssymb}
\usepackage{enumitem}
\usepackage{amsmath}
\usepackage{subfigure}
\usepackage{multirow}
\usepackage{enumitem}
\usepackage[ruled, vlined, boxed, linesnumbered]{algorithm2e}
\usepackage{bibentry}
\usepackage{adjustbox}
\usepackage[ruled,linesnumbered,vlined]{algorithm2e}
\makeatletter
\newcommand{\algorithmfootnote}[2][\footnotesize]{%
  \let\old@algocf@finish\@algocf@finish% Store algorithm finish macro
  \def\@algocf@finish{\old@algocf@finish% Update finish macro to insert "footnote"
    \leavevmode\rlap{\begin{minipage}{\linewidth}
    #1#2
    \end{minipage}}%
  }%
}
\makeatother
\usepackage{newfloat}
\usepackage{listings}
\DeclareCaptionStyle{ruled}{labelfont=normalfont,labelsep=colon,strut=off} % DO NOT CHANGE THIS
\usepackage{latexsym}
\usepackage{mathrsfs}
\usepackage{amsmath}
\usepackage{pifont}
\usepackage{caption}
\usepackage{inconsolata}
\usepackage{tabularx}
\usepackage{multirow}
\usepackage{amsthm}
\newtheorem{definition}{Definition}
\newcommand*\tcircle[1]{%
  \raisebox{-0.5pt}{%
    \textcircled{\fontsize{7pt}{0}\fontfamily{phv}\selectfont #1}%
  }%
}

\title{Hierarchical and Incremental Structural Entropy Minimization for Unsupervised Social Event Detection}
\author{
    Yuwei Cao\textsuperscript{\rm 1},
    Hao Peng\textsuperscript{\rm 2},
    Zhengtao Yu\textsuperscript{\rm 3},
    Philip S. Yu\textsuperscript{\rm 1}
}
\affiliations{
    \textsuperscript{\rm 1} Department of Computer Science, University of Illinois Chicago, Chicago, USA\\
    \textsuperscript{\rm 2} School of Cyber Science and Technology, Beihang University, Beijing, China\\
    \textsuperscript{\rm 3} Yunnan Key Laboratory of Artificial Intelligence, Kunming University of Science and Technology, Kunming, China\\
    \{ycao43, psyu\}@uic.edu, penghao@buaa.edu.cn, ztyu@hotmail.com
}

\usepackage{bibentry}
\begin{document}

\maketitle
\newcommand{\framework}{HISEvent}
\begin{abstract}
As a trending approach for social event detection, graph neural network (GNN)-based methods enable a fusion of natural language semantics and the complex social network structural information, thus showing SOTA performance. However, GNN-based methods can miss useful message correlations. Moreover, they require manual labeling for training and predetermining the number of events for prediction.
In this work, we address social event detection via graph structural entropy (SE) minimization.
While keeping the merits of the GNN-based methods, the proposed framework,~\framework, constructs more informative message graphs, is unsupervised, and does not require the number of events given a priori. Specifically, we incrementally explore the graph neighborhoods using 1-dimensional (1D) SE minimization to supplement the existing message graph with edges between semantically related messages. We then detect events from the message graph by hierarchically minimizing 2-dimensional (2D) SE. Our proposed 1D and 2D SE minimization algorithms are customized for social event detection and effectively tackle the efficiency problem of the existing SE minimization algorithms. Extensive experiments show that \framework~consistently outperforms GNN-based methods and achieves the new SOTA for social event detection under both closed- and open-set settings while being efficient and robust.
\end{abstract}

\begin{figure*}[t]
\centering
\includegraphics[width = 16cm]{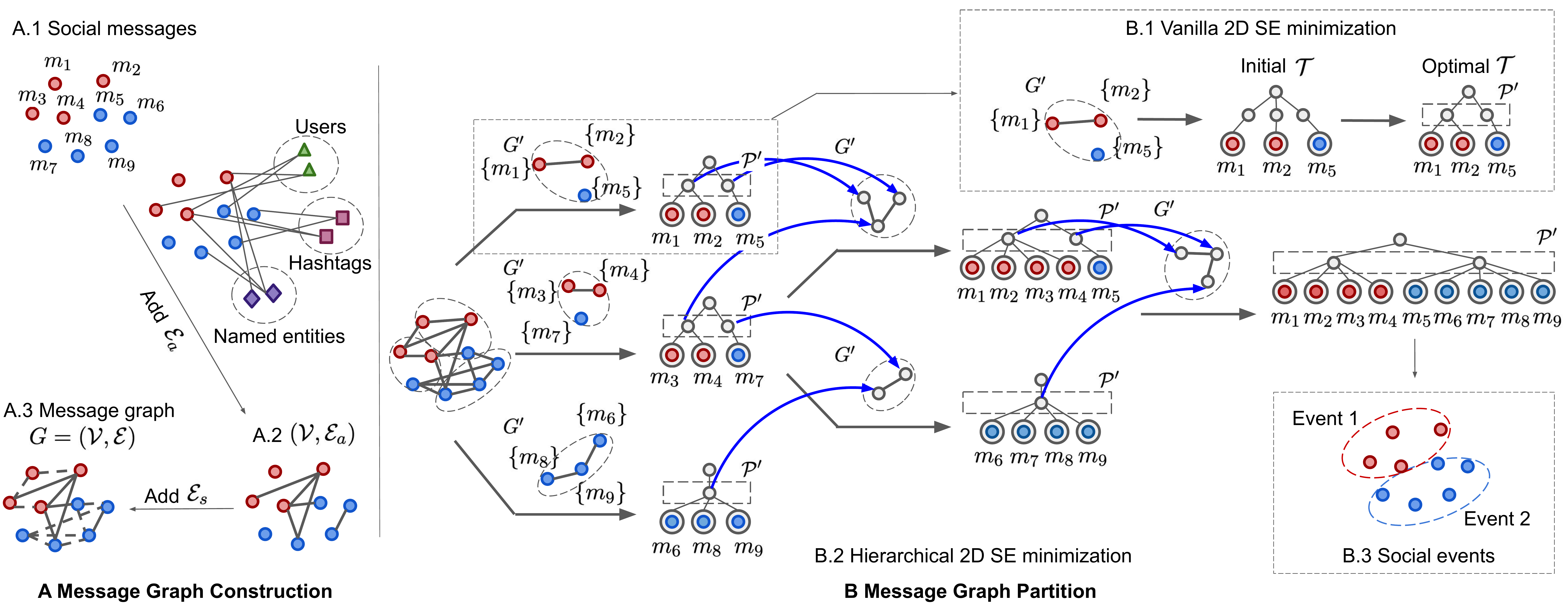}
\caption{The proposed \framework~framework. \textbf{A} and \textbf{B} are message graph construction and partitioning processes, respectively. An initial message graph \textbf{A.2} is constructed by linking social messages (\textbf{A.1}) that share common attributes. Further adding semantic-similarity-based edge set $\mathcal{E}_s$ results in the final message graph (\textbf{A.3}). \textbf{B.2} shows our proposed hierarchical 2D SE minimization algorithm, which repeatedly detects clusters ($\mathcal{P}'$) from sub-graphs ($G'$). \textbf{B.1} shows how clusters are detected in a single sub-graph via vanilla 2D SE minimization. \textbf{B.3} shows the detected social events.}
\label{fig:overall}
\end{figure*}

\section{Introduction}
Social event detection serves as a foundation for public opinion mining \cite{beck2021investigating}, fake news detection \cite{mehta2022tackling}, etc., and is attracting increasing attention in industry and academia.
%Social event detection is essential for crisis management, public opinion analysis, product recommendation, information retrieval, etc., and is attracting increasing attention in industry and academia \cite{cao2021knowledge}.
Existing studies \cite{ren2022known, cao2021knowledge, liu2020story, peng2019fine, peng2022reinforced} commonly formalize the task of social event detection as extracting clusters of co-related messages from sequences of social media messages.

Recent years have witnessed the booming of social event detection studies \cite{ren2023uncertainty, ren2022known, peng2022reinforced, cao2021knowledge, peng2019fine} that are based on Graph Neural Networks (GNN) \cite{kipf2016semi, velivckovic2017graph, hamilton2017inductive}. 
These methods typically follow a two-step strategy: they first construct message graphs that contain all the candidate messages, with ones that share common attributes (user mentions, hashtags, named entities, etc.) linked together.
Figure 1A.2 shows an example message graph. 
They then partition the message graph using GNNs, which incorporate the natural language representations of the messages with that of their neighbors. 
The resulting graph partitions (e.g., Figure 1B.3) serve as the detected social events.
Despite their SOTA performance, GNN-based methods merely link messages that share exactly the same attributes. 
The useful correlations between messages that are semantically close yet have no common attributes are missing. 
Furthermore, the GNN components of these models require supervision for training and predetermining the total number of events for prediction. Recent GNN-based methods \cite{ren2022known, peng2022reinforced, cao2021knowledge}, unlike earlier ones \cite{peng2019fine}, adopt contrastive learning, inductive learning, and pseudo label generation to alleviate the reliance on labels. 
However, manual labeling is still necessary for the initial training and periodical maintenance.

%hard match/ soft match.
In this work, we address the above issues from an information-theoretic perspective. 
We gain inspiration from \textit{structural entropy} (SE) \cite{li2016structural}, a metric that assesses the amount of information contained in a graph. 
Specifically, minimizing one-dimensional (1D) SE discloses the reliable node correlations contained in the raw, noisy graphs and is applied in biomedical studies \cite{li2016three}. 
We explore message graph neighborhoods via 1D SE minimization and supplement the existing message graph with edges between the semantically close messages. 
Unlike previous studies \cite{li2016three,li2018decoding}, our exploration is conducted in an incremental manner to maximize efficiency. 
Minimizing higher-dimensional SE decrypts the higher-order structure of the graphs \cite{li2016structural}. 
Given this, we further partition the message graph via two-dimensional (2D) SE minimization. 
Though effective and requiring no supervision, 2D SE minimization can be prohibitively slow to perform on complex, large-scale message graphs. 
We effectively tackle this by customizing a 2D SE minimization algorithm for social event detection. Our algorithm addresses the message correlations in a hierarchical manner: it repeatedly splits the message graph, detects clusters, and combines the clusters into new ones while keeping the previously detected partitions. 
Our proposed framework, hierarchical and incremental structural entropy minimization-guided social event detector (\framework), holds the merits of the GNN-based methods, learns more informative message graphs, and does not require supervision or the number of events given a priori. 
%We demonstrate that~\framework~scales to real-world monitoring applications such as emergency response after disasters and public opinion monitoring, which can intrigue rescue workers, government personnel, and media practitioners. We also show that~\framework~can be easily extended to streaming senarios.
Experiments on two public Twitter datasets show that~\framework~consistently outperforms strong baselines under both closed- and open-set settings and is the new SOTA for social event detection.
We also empirically show the efficiency and robustness of~\framework~as well as the effectiveness of all its components. 
%Our source code and pre-processed data are publicly available\footnote{\url{https://anonymous.4open.science/r/HISEvent-FD20}}.
Our contributions are: 
\begin{itemize}
\item We address social event detection from an information-theoretic lens. Compared to the GNN-based methods, the proposed \framework~learns more informative message graphs and requires no labeled samples or a predetermined number of events. 
To the best of our knowledge, we are the first to apply SE minimization for social event detection. 
\item We design novel SE minimization algorithms for social event detection. Besides being effective, \framework~efficiently runs on complex, large-scale message graphs. \framework~incrementally and hierarchically minimizes 1D and 2D SE, significantly reducing time complexity compared to the existing SE minimization algorithms. 
\item We conduct extensive experiments on two large, public Twitter datasets to show the new SOTA performance, efficiency, and robustness of \framework. Our code is publicly available \footnote{\url{https://github.com/SELGroup/HISEvent}}.
\end{itemize}
%\item We design novel SE minimization algorithms for social event detection. \framework~minimizes 1D and 2D SEs in incremental and hierarchical manners. Besides being effective, \framework~efficiently runs on complex, large-scale message graphs. 

\section{Preliminary}
Structural entropy (SE) \cite{li2016structural} is defined as the minimum number of bits to encode the vertex that is accessible with a step of random walk on a graph. 
The SE of a graph measures the complexity of the underlying essential structure and corresponds to an \textit{encoding tree}. 
SE can be of different dimensions, which measure the structural information of different orders and correspond to encoding trees of different heights. We present the formal definitions of encoding tree and SE as follows. Notations used in this paper are summarized in Appendix.
\begin{definition}
  \cite{li2016structural}. The encoding tree $\mathcal{T}$ of a graph $G=(\mathcal{V},\mathcal{E})$ is a hierarchical partition of $G$. 
It is a tree that satisfies the following: 
\begin{description}
\item 1) Each node $\alpha$ in $\mathcal{T}$ is associated with a set $T_\alpha \subseteq \mathcal{V}$. For the root node $\lambda$ of $\mathcal{T}$, $T_\lambda = \mathcal{V}$. Any leaf node $\gamma$ in $\mathcal{T}$ is associated with a single node in $G$, i.e., $T_\gamma = \{v\}, v \in \mathcal{V}$.
\item 2) For each node $\alpha$ in $\mathcal{T}$, denote all its children as $\beta_1, ..., \beta_k$, then $(T_{\beta_1}, ..., T_{\beta_k})$ is a partition of $T_\alpha$.
\item 3) For each node $\alpha$ in $\mathcal{T}$, denote its height as $h(\alpha)$. Let $h(\gamma) = 0$ and $h(\alpha^-) = h(\alpha) + 1$, where $\alpha^-$ is the parent of $\alpha$. The height of $\mathcal{T}$, $h(\mathcal{T}) = \underset{\alpha\in\mathcal{T}}{\max}\{h(\alpha)\}$.
\end{description}
\end{definition}

\begin{definition}
\label{definition:SE_T}
\cite{li2016structural}. The structural entropy (SE) of graph $G$ on encoding tree $\mathcal{T}$ is defined as:
\begin{equation}
\mathcal{H}^\mathcal{T}(G) = -\underset{\alpha\in\mathcal{T}, \alpha\neq\lambda}{\sum}\frac{g_\alpha}{vol(\lambda)}log\frac{vol(\alpha)}{vol(\alpha^-)},
\end{equation}
where $g_\alpha$ is the summation of the degrees (weights) of the cut edges of $T_\alpha$ (edges in $\mathcal{E}$ that have exactly one endpoint in $T_\alpha$). $vol(\alpha)$, $vol(\alpha^-)$, and $vol(\lambda)$ refer to the volumes, i.e., summations of the degrees of all the nodes, of $T_\alpha$, $T_{\alpha^-}$, and $T_\lambda$, respectively.
\end{definition}
%Based on the above definition, $\mathcal{H}^{(k)}(G) = \underset{\forall\mathcal{T}:h(\mathcal{T})=k}{\min}\{\mathcal{H}^{\mathcal{T}}(G)\}$ defines the \textit{$k$-dimensional SE} of $G$ and is realized by acquiring an optimal encoding tree of height $k$, in which the disturbance derived from noise or stochastic variation is minimized. Figure \ref{fig:overall} B.1 shows a toy example that constructs and optimizes $\mathcal{T}$ given $G'$.
The $d$-dimensional SE of $G$, defined as $\mathcal{H}^{(d)}(G) = \underset{\forall\mathcal{T}:h(\mathcal{T})=d}{\min}\{\mathcal{H}^{\mathcal{T}}(G)\}$, is realized by acquiring an optimal encoding tree of height $d$, in which the disturbance derived from noise or stochastic variation is minimized. Figure 1B.1 shows a toy example that constructs and optimizes $\mathcal{T}$ given $G'$.

\section{Methodology}
Figure 1 shows an overview of \framework. 
Following the previous methods \cite{ren2022known}, we adopt a two-step, message graph construction-partitioning strategy. 
We first formalize the task. Next, we propose to incorporate a novel semantic-similarity-based approach for message graph construction. We then present our unsupervised message graph partitioning. Finally, we analyze the time complexity. \framework, as a batched (retrospective) method, can be easily extended to streaming scenarios (discussed in Appendix). 
%\autoref{sec:problem_definition} formalizes our task. For message graph construction, we incorporate the existing common-attribute-based approach with a novel semantic-similarity-based approach, as discussed in \autoref{sec:1D_SE}. Our message graph partitioning, as presented in \autoref{sec:2D_SE}, is unsupervised and conducted from an information-theoretic perspective, without relying on GNNs or other deep-learning models. \autoref{sec:time} analyzes the time complexity of~\framework~and demonstrate that~\framework~scales to real-world applications. Finally, \autoref{sec:extend} shows that~\framework, as a batched (retrospective) social event detection method, can also be easily extended to streaming scenarios.
%\autoref
\subsection{Problem Formalization}
\label{sec:problem_definition}
Given a sequence of social messages $m_1, ..., m_N$ as input, the task of social event detection can be fulfilled by constructing and partitioning a message graph $G = (\mathcal{V},\mathcal{E})$. 
The node set $\mathcal{V} = \{m_1, ..., m_N\}$. 
The edge set $\mathcal{E}$ is initially empty and to be expanded by the message graph construction process. 
Partitioning $G$ results in $\{e_1, ..., e_M\}, e_i \subset \mathcal{V}, e_i \cap e_j = \emptyset$, which is a partition of $\mathcal{V}$ containing $M$ clusters (sets) of messages that correspond to the $M$ detected social events.
%e_i = \{m_{ij}|1\leq j \leq|e_i|\}

\subsection{Message Graph Construction with Incremental 1D SE Minimization}
\label{sec:1D_SE}
\setlength{\textfloatsep}{0cm}

\begin{algorithm}
\small
\caption{Determine $\mathcal{E_{\text{s}}}$ via incremental 1D SE minimization.}\label{algorithm:incremental_1D_SE}
\KwIn{Message graph node set $\mathcal{V}$}
%, which is the node set of $G$
\KwOut{Semantic-similarity-based edge set $\mathcal{E_{\text{s}}}$}
SEs $\gets \emptyset$\\
%\Init{$\mathcal{E} \gets \emptyset$, $k \gets 1$}\\
%\emph{// Construct $k$-NN edge sets}\\
Embed $\mathcal{V}$ via PLM and get $\{\boldsymbol{h}_{m_i}\}^{|\mathcal{V}|}_{i=1}$\\
%{\boldsymbol{h}_{m_i}|1\leq i \leq |\mathcal{V}|\}
%Calculate the correlation matrix
%For each message, sort all the rest messages 
\For(\tcp*[h]{Sort neighbors}){$i = 1, ..., |\mathcal{V}|$}{$\text{neighb}_{m_i} = (m_j)_{j=1}^{|\mathcal{V}|}$ s.t. $j\neq i$ and $\text{Cos}(\boldsymbol{h}_{m_i}, \boldsymbol{h}_{m_{j-1}}) > \text{Cos}(\boldsymbol{h}_{m_i}, \boldsymbol{h}_{m_{j}})$}
$\mathcal{E} \gets \{1\text{st element in } \text{neighb}_{m_i}\}^{|\mathcal{V}|}_{i=1}$\\
Calculate $\mathcal{H}^{(1)}(G)$ via Eq. 2\\
Append $\mathcal{H}^{(1)}(G)$ to SEs\\
$k = 2$\\
\While(\tcp*[h]{Search for the 1st stable point}){$k<|\mathcal{V}|$}{
$\mathcal{E} = \mathcal{E}\cup\{k\text{-th element in } \text{neighb}_{m_i}\}^{|\mathcal{V}|}_{i=1}$\\
Calculate $\mathcal{H}^{(1)\prime}(G)$ via Eq. 3\\
Append $\mathcal{H}^{(1)\prime}(G)$ to SEs\\
\If{$(k-1)$ is a stable point*}{break} 
$k = k+1$
}
$\mathcal{E_{\text{s}}} \gets \{(m_i, m_j)| m_j \in \text{the first (k-1) elements in } \text{neighb}_{m_i}\}^{|\mathcal{V}|}_{i=1}$\\
\Return{$\mathcal{E_{\text{s}}}$}
\algorithmfootnote{*$(k-1)$ is a stable point if the $(k-1)$-th element in SEs is smaller than the elements before and after it.}
\end{algorithm}
\setlength{\floatsep}{0cm}

Ideally, the edges in the message graph should faithfully reflect the reliable message correlations while eliminating the noisy ones. Following the GNN-based studies \cite{ren2022known, cao2021knowledge}, we capture the common-attribute-based message correlations, visualized in Figure 1A. Specifically, for each message $m_i$, we extract its attributes $A_i = \{u_i\}\cup\{um_{i_1}, um_{i_2}, ...\}\cup\{h_{i_1}, h_{i_2}, ...\}\cup\{ne_{i_1}, ne_{i_2}, ...\}$, where the RHS refers to a union of the sender, mentioned users, hashtags, and named entities associated with $m_i$. We add an edge $(m_i, m_j)$ into $\mathcal{E_{\text{a}}}$ iif $m_i$ and $m_j$ share some common attributes, i.e., $\mathcal{E_{\text{a}}} = \{(m_i, m_j)|A_i\cap A_j \neq \emptyset\}$.
%we first preprocess the data by filtering out URLs, extra characters, emotion icons, user IDs, etc., which we believe don’t have clear natural language semantics, from the message contents to learn message embeddings.

$\mathcal{E_{\text{a}}}$ alone, however, can miss useful correlations, as there are messages that have similar semantics yet share no common attributes. 
To mitigate this, we supplement the message graph with semantic-similarity-based edges, denoted as $\mathcal{E}_s$.
The similarity between two messages can be measured by embedding \footnote{Before embedding, we preprocess the message contents by filtering out URLs, extra characters, emotion icons, and user IDs, which we believe don’t have clear natural language semantics.} them via pre-trained language models (PLMs), i.e., SBERT \cite{reimers2019sentence} then calculating the cosine similarity between their representations. 
The idea is to link each message to its $k$-nearest neighbors, where $k$ needs to be carefully chosen to keep only the reliable connections. 
1D SE minimization has been applied in biomedical studies \cite{li2016three} to select the most correlated neighbors.
%, i.e., determine $k$. 
Nonetheless, \cite{li2016three} calculates the 1D SE from scratch for every candidate $k$, which is inefficient. 
We propose incremental 1D SE minimization for correlated neighbor selection.
Specifically, we start with $\mathcal{E}_s = \emptyset$ and incrementally insert sets of edges into $G$, with the $k$-th set (referred to as $k$-NN edge set) containing edges between each node and its $k$-th nearest neighbor. 
The initial 1D SE with $k=1$ is:
\begin{equation}
\label{eq:initial_1D}
\mathcal{H}^{(1)}(G) = -\sum_{i=1}^{|\mathcal{V}|}\frac{d_i}{vol(\lambda)}log\frac{d_i}{vol(\lambda)},
\end{equation}
and the successive updates follow:
\begin{equation}
\label{eq:increment_1D}
%\begin{multline}
\begin{split}
& \mathcal{H}^{(1)\prime}(G) = \frac{vol(\lambda)}{ vol^\prime(\lambda)}\Bigl(\mathcal{H}^{(1)}(G)-log\frac{vol(\lambda)}{vol^\prime(\lambda)}\Bigr) \\
& + \sum_{j=1}^{|a_k|}\Bigl(\frac{d_j}{vol^\prime(\lambda)}log\frac{d_j}{vol^\prime(\lambda)} -\frac{d_j^\prime}{vol^\prime(\lambda)}log\frac{d_j^\prime}{vol^\prime(\lambda)}\Bigr),
\end{split}
\end{equation}
%where $d_i$ denotes the original degree (weighted) of a node $i$ in $G$ (initially, $d_i$ is calculated with $i$ linking to its 1st nearest neighbor). $d_i^\prime$ denotes the updated degree of $i$ with an edge between $i$ and its $k$-th nearest neighbor inserted into $G$. 
where $d_i$ and $d_i^\prime$ denote the original and updated degrees (weighted) of node $i$ in $G$ before and after the insertion of the $k$-NN edge set, respectively. Initially, $d_i$ is calculated with $i$ linking to its 1st nearest neighbor. $a_k$ is a set of nodes whose degrees are affected by the insertion of the $k$-NN edge set. $vol(\lambda)$ and $vol^\prime(\lambda)$ stand for the volumes of $G$ before and after inserting the $k$-NN edge set. $\mathcal{H}^{(1)}(G)$ and $\mathcal{H}^{(1)\prime}(G)$ stand for the original and updated 1D SE. The derivation of Equation 3 is in Appendix. 

With the above initialization and update rules, selecting the proper $k$ then follows Algorithm 1. 
Compared to \cite{li2016three}, the time needed for inspecting each candidate $k$ (lines 10-12) is reduced from $O(|\mathcal{V}|)$ to $O(|a_k|)$ ($|a_k| \leq |\mathcal{V}|$ always holds). 
Another difference is, \framework~only uses $\mathcal{E_{\text{s}}}$ as a supplementation to $\mathcal{E_{\text{a}}}$. We, therefore, adopt the first stable point (lines 13-14) instead of the global one. The overall running time, due to lines 3-4, is $O(|\mathcal{V}|^2)$.

%Finally, letting $\mathcal{E} = \mathcal{E_{\text{a}}} \cup \mathcal{E_{\text{s}}}$ and adding weight $w_{ij}=max(\text{Cos}(\boldsymbol{h}_{m_i}, \boldsymbol{h}_{m_{j}}), 0)$ for each edge $(m_i, m_j)$ accomplish the construction of the message graph. 
Finally, we set $\mathcal{E} = \mathcal{E_{\text{a}}} \cup \mathcal{E_{\text{s}}}$. For each edge $(m_i, m_j)$, we then set its weight $w_{ij}=max(\textit{cosine}(\boldsymbol{h}_{m_i}, \boldsymbol{h}_{m_{j}}), 0)$, where $\boldsymbol{h}_{m_{i}}$ and $\boldsymbol{h}_{m_{j}}$ denote the embeddings of $m_i$ and $m_j$ learned via PLMs. This accomplishes the construction of the message graph. \framework~incorporates not only the common-attributes-based message correlations but also the semantic-similarity-based ones. It constructs more informative message graphs compared to the previous studies \cite{ren2022known, peng2022reinforced, cao2021knowledge}.

%Please note that~\framework~is a pioneering work that verifies, on single-relation message graphs, the idea of supplementing natural language semantics with structural information in an unsupervised manner for efficient and effective social event detection.~\framework~can be extended to consider multiple relationships by weighting the edges in the message graph differently, according to their types. One naive approach would be treating the edge-type weights as hyperparameters to be decided empirically. We recognize that this may further improve the performance but leave the exploration to the future. 

\subsection{Event Detection via Hierarchical 2D SE Minimization}
\label{sec:2D_SE}
Message graph partitioning decodes $G$ into $\mathcal{P}$, which contains the detected events in the form of message clusters.
A faithful decoding of the message correlations in $G$ assigns related messages to the same cluster and unrelated ones to different clusters. 
%$\mathcal{P}$ should faithfully decode the message correlations in $G$ by assigning the related messages to the same cluster and the unrelated ones to different clusters. 
Previous GNN-based detectors \cite{ren2022known, cao2021knowledge} learn to properly partition message graphs through training, which require costly sample labeling and the number of events a priori. 
To address this issue, \framework~conducts unsupervised partitioning under the guidance of 2D SE minimization, which eliminates the noise and reveals the essential 2nd-order (cluster-wise) structure underneath the raw graph with no prior knowledge of the number of event clusters. 

\cite{li2016structural} proposes a vanilla greedy 2D SE minimization algorithm that repeatedly merges any two nodes in the encoding tree $\mathcal{T}$ that would result in the largest decrease in 2D SE until reaches the minimum possible value. Hence it partitions a graph without supervision or a predetermined total number of clusters. We illustrate this algorithm in Appendix. This vanilla 2D SE minimization, however, takes $O({|\mathcal{V}|}^3)$ to run. 
Though works for small bio-informatics graphs \cite{wu2022structural}, it is prohibitively slow for the large, complex message graphs (demonstrated by Section 4.4). 
To address this, we propose to minimize 2D SE and detect events in a hierarchical manner, shown in Algorithm 2.
Specifically, each message is initially in its own cluster (line 1). 
We split the clusters into subsets of size $n$ (line 3) and merge the clusters involved in each subset using the vanilla greedy algorithm to get new clusters (lines 5-13). 
The new clusters are then passed on to the next iteration (line 14). 
This process is repeated until the clusters that contain all the messages are considered simultaneously (lines 15-16). 
If, at some point, none of the clusters in any subset can be merged, we increase $n$ so that more clusters can be considered in the same subset and, therefore, may be merged (lines 17-18).
%Figure \ref{fig:overall}B visualizes this process with an example. Note how $n=3$ clusters are considered at a time, and the partitions determined by the previous iteration are always passed on to the later iteration.
Figure 1B visualizes this process: $m_1$ to $m_9$ are initially in their own clusters. $n=3$ clusters are considered at a time to form a $G'$. Clusters in each $G'$ are then merged via vanilla 2D SE minimization to get $\mathcal{P}'$ (Figure 1B.1). The partitions resulted in the previous iteration are passed on to the later iteration, as indicated by the blue curved arrows in Figure 1B.2. The process terminates when a $\mathcal{P}'$ that involves all the messages is determined. With a running time of $O(n^3)$, Algorithm 2 is much more efficient than its vanilla predecessor, as $n$ is a hyperparameter that can be set to $\ll |\mathcal{V}|$. To summarize, \framework~detects social events from the complex message graphs in an effective and unsupervised manner.

\begin{algorithm}
\small
\caption{Event detection via hierarchical 2D SE minimization.}\label{algorithm:hierarchical_2D_SE}
\KwIn{Message graph $G = (\mathcal{V}$, $\mathcal{E})$, sub-graph size $n$}
\KwOut{A partition $\mathcal{P}$ of $\mathcal{V}$}
$\mathcal{P} \gets (m|m \in \mathcal{V})$

\While{True}{
    $\{\mathcal{P}_{s}\} \gets $ consecutively remove the first $min(n, \text{size of the remaining part of } \mathcal{P})$ clusters from $\mathcal{P}$ that form a set $\mathcal{P}_{s}$
    
    \For{$\mathcal{P}_{s} \in \{\mathcal{P}_{s}\}$}{
        $\mathcal{V}^{'} \gets \text{combine all the clusters in } \mathcal{P}_{s}$
        
        $\mathcal{E}^{'} \gets \{e \in \mathcal{E}, \text{both endpoints of }e \in \mathcal{V}^{'}\}$
        
        ${G'} \gets (\mathcal{V}', \mathcal{E}')$
        
        $\mathcal{T}' \gets$ add a root tree node $\lambda$
        
        \For{cluster $\mathcal{C} \in \mathcal{P}_{s}$}{
            Add a tree node $\alpha$ to $\mathcal{T}'$, s.t. $\alpha^- = \lambda, T_{\alpha} = \mathcal{C}$
            
            \For{message $m \in \mathcal{C}$}{
                Add a tree node $\gamma$ to $\mathcal{T}'$, s.t. $\gamma^- = \alpha, T_{\gamma} = \{m\}$
            }
        }
        $\mathcal{P}' \gets$ run vanilla 2D SE minimization (see Appendix) on $G'$, with the initial encoding tree set to $\mathcal{T}'$
        
        Append $\mathcal{P}'$ to $\mathcal{P}$
    }
    \If{$|\{\mathcal{V}^{'}\}| = 1$}{
        Break
    }
    \If{$\mathcal{P}$ is the same as at the end of last iteration}{
        $n \gets 2n$
    }
}
\Return $\mathcal{P}$
\end{algorithm}

\subsection{Time Complexity of~\framework}
\label{sec:time}
The overall time complexity of \framework~is $O(|\mathcal{E}_a|+|\mathcal{V}|^2+n^3)$, where $|\mathcal{E}_a|$ is the total number of common-attribute-based edges in the message graph, $|\mathcal{V}|$ is the total number of nodes (i.e., messages) and $n$ is sub-graph size, a hyperparameter that can be set to $\ll |\mathcal{V}|$. 
Specifically, the running time of constructing $\mathcal{E}_a$ is $O(|\mathcal{E}_a|)$. 
The running time of constructing the semantic-similarity-based edge set $\mathcal{E}_s$ is $O(|\mathcal{V}|^2)$. 
The running time of detecting social events from the constructed message graph is $O(n^3)$. 
Note~\framework~can be easily parallelized (discussed in Appendix).

\begin{table*}[t]
\centering
\begin{adjustbox}{width=0.8\linewidth}
    \begin{tabular}{c|c|ccccc|cc}
    \hline
    Dataset & Metric & KPGNN* & QSGNN* & EventX & BERT* & SBERT* & \framework & Improv. (\%)\\
    \hline
    \multirow{2}{*}{Event2012} & ARI & \textit{0.22} & \textit{0.22} & 0.05 & 0.12 & 0.17 & \textbf{0.50} & $\uparrow$127\\
    & AMI & 0.52 & 0.53 & 0.19 & 0.43 & \textit{0.73} & \textbf{0.81} & $\uparrow$11 \\
    \hline
    \multirow{2}{*}{Event2018} & ARI & 0.15 & \textit{0.16} & 0.03 & 0.05 & 0.11 & \textbf{0.44} & $\uparrow$175 \\
    & AMI & 0.44 & 0.44 & 0.16 & 0.34 & \textit{0.62} & \textbf{0.66} & $\uparrow$6 \\
    \hline
    \end{tabular}
    \end{adjustbox}
    \caption{Closed-set results. * marks results acquired with the ground truth event numbers.}
  \label{table:closeset_results}
\end{table*}

\begin{table*}[t]
  \centering
  \begin{adjustbox}{width=0.9\linewidth}
    \begin{tabular}{c|cc|cc|cc|cc|cc|cc|cc}
    \hline
    Blocks ($\#$events) & \multicolumn{2}{c}{$M_1$ (41)} & \multicolumn{2}{c}{$M_2$ (30)} & \multicolumn{2}{c}{$M_3$ (33)} & \multicolumn{2}{c}{$M_4$ (38)} & \multicolumn{2}{c}{$M_5$ (30)} & \multicolumn{2}{c}{$M_6$ (44)} & \multicolumn{2}{c}{$M_7$ (57)} \\
    \hline
    Metric & ARI & AMI & ARI & AMI & ARI & AMI & ARI & AMI & ARI & AMI & ARI & AMI & ARI & AMI \\
    \hline
    KPGNN* & \textit{0.07} & 0.37 & 0.76 & 0.78 & 0.58 & 0.74 & 0.29 & 0.64 & 0.47 & 0.71 & 0.72 & 0.79 & \textit{0.12} & 0.51 \\ 
    QSGNN* & \textit{0.07} & \textit{0.41} & \textit{0.77} & 0.80 & 0.59 & 0.76 & 0.29 & 0.68 & 0.48 & \textit{0.73} & \textit{0.73} & 0.80 & \textit{0.12} & 0.54 \\
    EventX & 0.01 & 0.06 & 0.45 & 0.29 & 0.09 & 0.18 & 0.07 & 0.19 & 0.04 & 0.14 & 0.14 & 0.27 & 0.02 & 0.13 \\
    BERT* &  0.03 & 0.35 & 0.65 & 0.76 & 0.45 & 0.72 & 0.19 & 0.58 & 0.36 & 0.67 & 0.45 & 0.75 & 0.07 & 0.50 \\
    SBERT* & 0.03 & 0.38 & 0.73 & \textit{0.85} & \textit{0.68} & \textit{0.87} & \textit{0.36} & \textit{0.80} & \textit{0.61} & \textbf{0.85} & 0.53 & \textit{0.83} & 0.09 & \textit{0.61} \\
    \hline
    \framework & \textbf{0.08} & \textbf{0.44} & \textbf{0.79} & \textbf{0.88} & \textbf{0.95} & \textbf{0.94} & \textbf{0.50} & \textbf{0.84} & \textbf{0.62} & \textbf{0.85} & \textbf{0.86} & \textbf{0.90} & \textbf{0.27} & \textbf{0.68} \\
    Improv. (\%) & $\uparrow$14 & $\uparrow$7 & $\uparrow$3 & $\uparrow$4 & $\uparrow$40 & $\uparrow$8 & $\uparrow$39 & $\uparrow$5 & $\uparrow$2 & $\rightarrow$ & $\uparrow$18 & $\uparrow$8 & $\uparrow$125 & $\uparrow$11 \\
    \hline
    \hline
    Blocks ($\#$events) & \multicolumn{2}{c}{$M_8$ (53)} & \multicolumn{2}{c}{$M_9$ (38)} & \multicolumn{2}{c}{$M_{10}$ (33)} & \multicolumn{2}{c}{$M_{11}$ (30)} & \multicolumn{2}{c}{$M_{12}$ (42)} & \multicolumn{2}{c}{$M_{13}$ (40)} & \multicolumn{2}{c}{$M_{14}$ (43)} \\
    \hline
    Metric & ARI & AMI & ARI & AMI & ARI & AMI & ARI & AMI & ARI & AMI & ARI & AMI & ARI & AMI \\
    \hline
    KPGNN* & 0.60 & 0.76 & 0.46 & 0.71 & 0.70 & 0.78 & \textit{0.49} & 0.71 & 0.48 & 0.66 & \textit{0.29} & 0.67 & \textit{0.42} & 0.65 \\ 
    QSGNN* & 0.59 & 0.75 & \textit{0.47} & 0.75 & \textit{0.71} & 0.80 & \textit{0.49} & \textit{0.72} & 0.49 & 0.68 & \textit{0.29} & 0.66 & 0.41 & 0.66 \\
    EventX & 0.09 & 0.21 & 0.07 & 0.19 & 0.13 & 0.24 & 0.16 & 0.24 & 0.07 & 0.16 & 0.04 & 0.16 & 0.10 & 0.14 \\
    BERT* & 0.51 & 0.74 & 0.34 & 0.71 & 0.55 & 0.78 & 0.26 & 0.62 & 0.31 & 0.56 & 0.13 & 0.57 & 0.24 & 0.55 \\
    SBERT* & \textit{0.65} & \textit{0.86} & \textit{0.47} & \textit{0.83} & 0.62 & \textit{0.85} & \textit{0.49} & \textbf{0.82} & \textit{0.63} & \textit{0.85} & 0.24 & \textit{0.70} & 0.40 & \textit{0.77} \\
    \hline
    \framework & \textbf{0.74} & \textbf{0.89} & \textbf{0.65} & \textbf{0.88} & \textbf{0.87} & \textbf{0.90} & \textbf{0.62} & \textbf{0.82} & \textbf{0.82} & \textbf{0.90} & \textbf{0.46} & \textbf{0.78} & \textbf{0.85} & \textbf{0.88} \\
    Improv. (\%) & $\uparrow$14 & $\uparrow$3 & $\uparrow$38 & $\uparrow$6 & $\uparrow$23 & $\uparrow$6 & $\uparrow$27 & $\rightarrow$ & $\uparrow$30 & $\uparrow$6 & $\uparrow$59 & $\uparrow$11 & $\uparrow$102 & $\uparrow$14 \\
    \hline
    \hline
    Blocks ($\#$events) & \multicolumn{2}{c}{$M_{15}$ (42)} & \multicolumn{2}{c}{$M_{16}$ (27)} & \multicolumn{2}{c}{$M_{17}$ (35)} & \multicolumn{2}{c}{$M_{18}$ (32)} & \multicolumn{2}{c}{$M_{19}$ (28)} & \multicolumn{2}{c}{$M_{20}$ (34)} & \multicolumn{2}{c}{$M_{21}$ (32)} \\
    \hline
    Metric & ARI & AMI & ARI & AMI & ARI & AMI & ARI & AMI & ARI & AMI & ARI & AMI & ARI & AMI \\
    \hline
    KPGNN* & \textit{0.17} & 0.54 & \textit{0.66} & 0.77 & 0.43 & 0.68 & 0.47 & 0.66 & 0.51 & 0.71 & 0.51 & 0.68 & 0.20 & 0.57\\ 
    QSGNN* & \textit{0.17} & 0.55 & 0.65 & 0.76 & \textit{0.44} & 0.69 & 0.48 & 0.68 & 0.50 & 0.70 & 0.51 & 0.69 & 0.21 & 0.58 \\
    EventX & 0.01 & 0.07 & 0.08 & 0.19 & 0.12 & 0.18 & 0.08 & 0.16 & 0.07 & 0.16 & 0.11 & 0.18 & 0.01 & 0.10 \\
    BERT* & 0.07 & 0.43 & 0.43 & 0.71 & 0.22 & 0.56 & 0.24 & 0.52 & 0.28 & 0.59 & 0.32 & 0.60 & 0.17 & 0.54 \\
    SBERT* & \textit{0.17} & \textit{0.67} & 0.50 & \textit{0.78} & 0.35 & \textit{0.77} & \textit{0.52} & \textbf{0.81} & \textit{0.54} & \textit{0.83} & \textit{0.52} & \textit{0.80} & \textit{0.24} & \textbf{0.70} \\
    \hline
    \framework & \textbf{0.27} & \textbf{0.72} & \textbf{0.83} & \textbf{0.87} & \textbf{0.56} & \textbf{0.81} & \textbf{0.70} & \textit{0.80} & \textbf{0.63} & \textbf{0.87} & \textbf{0.69} & \textbf{0.81} & \textbf{0.45} & \textit{0.69} \\
    Improv. (\%) & $\uparrow$59 & $\uparrow$7 & $\uparrow$26 & $\uparrow$12 & $\uparrow$27 & $\uparrow$5 & $\uparrow$35 & $\downarrow$1 & $\uparrow$17 & $\uparrow$5 & $\uparrow$33 & $\uparrow$1 & $\uparrow$88 & $\downarrow$1 \\
    \hline
  \end{tabular}
  \end{adjustbox}
  \caption{Open-set results on Event2012. * marks results acquired with the ground truth event numbers.}
  \label{table:openset_results_event2012}
\end{table*}

\begin{table*}[t]
  \centering
  \begin{adjustbox}{width=0.9\linewidth}
    \begin{tabular}{c|cc|cc|cc|cc|cc|cc|cc|cc}
    \hline
    Blocks & \multicolumn{2}{c}{$M_1$} & \multicolumn{2}{c}{$M_2$} & \multicolumn{2}{c}{$M_3$} & \multicolumn{2}{c}{$M_4$} & \multicolumn{2}{c}{$M_5$} & \multicolumn{2}{c}{$M_6$} & \multicolumn{2}{c}{$M_7$} & \multicolumn{2}{c}{$M_8$} \\
    \hline
    Metric & ARI & AMI & ARI & AMI & ARI & AMI & ARI & AMI & ARI & AMI & ARI & AMI & ARI & AMI & ARI & AMI \\
    \hline
    KPGNN* & 0.17 & 0.54 & 0.18 & 0.55 & 0.15 & 0.55 & 0.17 & 0.55 & 0.21 & 0.57 & 0.21 & 0.57 & \textit{0.30} & 0.61 & 0.20 & 0.57 \\ 
    QSGNN* & 0.18 & 0.56 & 0.19 & 0.57 & 0.17 & 0.56 & 0.18 & 0.57 & 0.23 & 0.59 & 0.21 & 0.59 & \textit{0.30} & 0.63 & 0.19 & 0.55 \\
    EventX & 0.02 & 0.11 & 0.02 & 0.12 & 0.01 & 0.11 & 0.06 & 0.14 & 0.13 & 0.24 & 0.08 & 0.15 & 0.02 & 0.12 & 0.09 & 0.21 \\
    BERT* & 0.16 & 0.42 & 0.21 & 0.44 & 0.22 & 0.44 & 0.17 & 0.41 & 0.31 & 0.56 & 0.23 & 0.49 & 0.23 & 0.49 & 0.24 & 0.50 \\
    SBERT* & \textit{0.20} & \textit{0.60} & \textit{0.29} & \textit{0.61} & \textit{0.34} & \textit{0.63} & \textit{0.23} & \textit{0.60} & \textit{0.47} & \textit{0.76} & \textit{0.41} & \textit{0.73} & 0.29 & \textit{0.65} & \textit{0.50} & \textit{0.75} \\
    \hline
    \framework & \textbf{0.55} & \textbf{0.77} & \textbf{0.67} & \textbf{0.79} & \textbf{0.47} & \textbf{0.74} & \textbf{0.46} & \textbf{0.72} & \textbf{0.66} & \textbf{0.82} & \textbf{0.61} & \textbf{0.83} & \textbf{0.56} & \textbf{0.81} & \textbf{0.82} & \textbf{0.90} \\
    Improv. (\%) & $\uparrow$175 & $\uparrow$28 & $\uparrow$131 & $\uparrow$30 & $\uparrow$38 & $\uparrow$17 & $\uparrow$100 & $\uparrow$20 & $\uparrow$40 & $\uparrow$8 & $\uparrow$49 & $\uparrow$14 & $\uparrow$87 & $\uparrow$25 & $\uparrow$64 & $\uparrow$20 \\
    \hline
    \hline
    Blocks & \multicolumn{2}{c}{$M_9$} & \multicolumn{2}{c}{$M_{10}$} & \multicolumn{2}{c}{$M_{11}$} & \multicolumn{2}{c}{$M_{12}$} & \multicolumn{2}{c}{$M_{13}$} & \multicolumn{2}{c}{$M_{14}$} & \multicolumn{2}{c}{$M_{15}$} & \multicolumn{2}{c}{$M_{16}$} \\
    \hline
    Metric & ARI & AMI & ARI & AMI & ARI & AMI & ARI & AMI & ARI & AMI & ARI & AMI & ARI & AMI & ARI & AMI \\
    \hline
    KPGNN* & 0.10 & 0.46 & 0.18 & 0.56 & 0.16 & 0.53 & 0.17 & 0.56 & 0.28 & 0.60 & 0.43 & 0.65 & 0.25 & 0.58 & 0.13 & 0.50 \\ 
    QSGNN* & 0.13 & 0.46 & 0.19 & 0.58 & 0.20 & 0.59 & 0.20 & 0.59 & 0.27 & 0.58 & \textit{0.44} & 0.67 & 0.27 & 0.61 & 0.13 & 0.50 \\
    EventX & 0.07 & 0.16 & 0.07 & 0.19 & 0.06 & 0.18 & 0.09 & 0.20 & 0.06 & 0.15 & 0.11 & 0.22 & 0.11 & 0.22 & 0.01 & 0.10 \\
    BERT* & 0.17 & 0.42 & 0.19 & 0.46 & 0.18 & 0.48 & 0.32 & 0.54 & 0.18 & 0.40 & 0.27 & 0.52 & 0.28 & 0.53 & 0.21 & 0.43 \\
    SBERT* & \textit{0.23} & \textit{0.63} & \textit{0.39} & \textit{0.72} & \textit{0.31} & \textit{0.70} & \textit{0.54} & \textit{0.76} & \textit{0.34} & \textit{0.65} & 0.43 & \textit{0.68} & \textit{0.40} & \textit{0.71} & \textit{0.25} & \textit{0.65} \\
    \hline
    \framework & \textbf{0.65} & \textbf{0.73} & \textbf{0.51} & \textbf{0.80} & \textbf{0.44} & \textbf{0.79} & \textbf{0.86} & \textbf{0.88} & \textbf{0.83} & \textbf{0.89} & \textbf{0.80} & \textbf{0.89} & \textbf{0.70} & \textbf{0.84} & \textbf{0.37} & \textbf{0.73} \\
    Improv. (\%) & $\uparrow$183 & $\uparrow$16 & $\uparrow$31 & $\uparrow$11 & $\uparrow$42 & $\uparrow$13 & $\uparrow$59 & $\uparrow$16 & $\uparrow$144 & $\uparrow$37 & $\uparrow$82 & $\uparrow$31 & $\uparrow$75 & $\uparrow$18 & $\uparrow$48 & $\uparrow$12 \\
    \hline
  \end{tabular}
  \end{adjustbox}
  \caption{Open-set results on Event2018. * marks results acquired with the ground truth event numbers.} 
  \label{table:openset_results_event2018}
\end{table*}

\begin{table}[t]
  \centering
  \begin{adjustbox}{width=0.9\linewidth}
    \begin{tabular}{c|cc|cc}
    \hline
    Setting & \multicolumn{2}{c}{Closed-set} & \multicolumn{2}{c}{Open-set (Avg.)} \\
    \hline
    Metric & ARI & AMI & ARI & AMI \\
    \hline
    \framework & \textbf{0.50} & \textbf{0.81} & \textbf{0.63} & \textbf{0.82}\\
    $-\mathcal{E}_s$ & 0.24 & 0.58 & 0.40 & 0.60 \\
    $-\mathcal{E}_a$ & \textit{0.42} & \textit{0.80} & 0.51 & \textit{0.77} \\
    \framework-BERT & 0.25 & 0.65 & 0.52 & 0.69 \\
    \framework-vanilla & \multicolumn{2}{c|}{takes $>$ 10 days} & \textit{0.62} & \textbf{0.82} \\
    \hline
    \end{tabular}
    \end{adjustbox}
\caption{Ablation study on Event2012. $-\mathcal{E}_s$ removes the semantic-similarity-based $\mathcal{E}_s$ and simply relies on the common-attribute-based $\mathcal{E}_a$ to capture message correlations. 
Similarly, $-\mathcal{E}_a$ relies solely on $\mathcal{E}_s$ (unlike in Section 3.2, here we use the global rather than the first stable points since $\mathcal{E}_s$ is no longer a supplementation but aims to fully capture the message correlations). \framework-BERT uses BERT rather than SBERT to measure the edge weights. \framework-vanilla partitions the message graph via vanilla 2D SE minimization instead of our proposed hierarchical one.} 
\label{table:ablation_study}
\end{table}

\section{Experiments}
\label{sec:experiments}
%We evaluate \framework~with extensive experiments. \autoref{sec:experimental_setup} is the experimental settings. \autoref{sec:overall_performance} compares \framework~to various baselines. \autoref{sec:ablation_study} shows the effectiveness of \framework's components. \autoref{sec:efficiency} and \autoref{sec:hyper_parameter} analyze the efficiency and hyperparameter sensitivity of \framework, respectively. \autoref{sec:case_study} presents a case study.
We conduct extensive experiments to compare~\framework~to various baselines and show the effectiveness of its components. We further analyze the efficiency as well as hyperparameter sensitivity of~\framework~and present a case study.

\subsection{Experimental Setup}
\label{sec:experimental_setup}
\textbf{Datasets.} 
We experiment on two large, public Twitter datasets, i.e., Event2012 \cite{mcminn2013building}, and Event2018 \cite{mazoyer2020french}. 
Event2012 contains 68,841 English tweets related to 503 events, spreading over four weeks. 
Event2018 contains 64,516 French tweets about 257 events and were sent within a span of 23 days. We evaluate under both closed- and open-set settings by adopting the data splits of \citealt{ren2022known} and \citealt{cao2021knowledge}. 
The former simultaneously consider all the events, while the latter assumes the events happen over time and splits the datasets into day-wise message blocks (e.g., $M_1$ to $M_{21}$ in Event2012).
Data statistics are in Appendix.
%Statistics on the data splits are in Appendix \ref{sec:data_splits}.
%We evaluate under both closed- and open-set settings \cite{ren2022known}. The former simultaneously consider all the events, while the latter assumes the events happen over time and partitions the datasets into day-wise message blocks. Although the proposed \framework~is fully unsupervised, some of the baseline methods require training. We, therefore, follow the data splits as adopted by those baselines \cite{cao2021knowledge, ren2022known}. For the closed-set (offline) scenario, the two datasets are randomly split by 70:10:20 into training, validation, and test sets. For the open-set (online) scenario, the messages of the first week form an initial block ($M_0$) for training and validation, while the successive day-wise message blocks ($M_1$ through $M_{21}$ for Event2012 and $M_1$ through $M_{16}$ for Event2018) are used for testing. Tables \ref{table:Closeset_data_splits}, \ref{table:Openset_data_splits_Event2012}, and \ref{table:Openset_data_splits_Event2018} show the statistics of the data splits.

\noindent\textbf{Baselines.} 
We compare \framework~to \textbf{KPGNN} \cite{cao2021knowledge}, a GNN-based social event detector, \textbf{QSGNN} \cite{ren2022known}, which improves upon KPGNN using restricted pseudo labels and is the current SOTA, and \textbf{EventX} \cite{liu2020story}, a non-GNN-based social event detector leverages community detection.
We also experiment on PLMs, i.e., \textbf{BERT} \cite{devlin2018bert}, and \textbf{SBERT} \cite{reimers2019sentence}:
%, by directly applying K-means clustering on the message embeddings learned via them.
we first input the preprocessed message contents to PLMs to learn message embeddings and then apply K-means clustering on the message embeddings to acquire events, i.e., message clusters.
Note that KPGNN and QSGNN are supervised. KPGNN, QSGNN, BERT, and SBERT require the total number of events to be specified a priori, which is impractical. \framework, in contrast, is unsupervised and does not need the total number of events as an input. 
%Please also note that as a long-standing task, social event detection intrigued a large number of studies based on various techniques. Our baselines outperform various previous techniques, including the naive ones. E.g., KPGNN outperforms TF-IDF \cite{bafna2016document}, LDA \cite{blei2003latent}, WMD \cite{kusner2015word}, LSTM \cite{graves2005framewise}, word2vec \cite{mikolov2013efficient}, etc. EventX outperforms co-clustering \cite{dhillon2003information}, NMF \cite{xu2003document}, etc. Therefore, we omit the direct comparison with these previous techniques.
Also note we omit the direct comparison with various techniques that are outperformed by the baselines, i.e., TF-IDF \cite{bafna2016document}, LDA \cite{blei2003latent}, WMD \cite{kusner2015word}, LSTM \cite{graves2005framewise}, word2vec \cite{mikolov2013efficient}, co-clustering \cite{dhillon2003information}, NMF \cite{xu2003document}, etc. Implementation details are in Appendix.

%\noindent\textbf{Experiment Setting.} 
%For \framework, we adopt SBERT to calculate edge weights (the calculation follows \autoref{sec:1D_SE} and the effects of changing the PLM are observed in \autoref{sec:ablation_study}). We set the sub-graph size $n$ for the closed-set experiments to 300 and 800 for Event2012 and Event2018, respectively. As to the open-set experiments, we set $n$ to 400 and 300 for Event2012 and Event2018 (the effects of changing $n$ are observed in \autoref{sec:hyper_parameter}). 
%For KPGNN, QSGNN, and EventX, we adopt the settings as reported in the original papers. For BERT, we adopt Hugging Face\footnote{\url{https://huggingface.co/models?filter=bert}} pretrained models, i.e., 'bert-large-cased' for Event2012 and 'bert-base-multilingual-cased' for Event2018. We apply mean pooling (i.e., average the last hidden states of the words in a message for its embedding) as we empirically found it outperforms the pooler and [CLS] output. For SBERT, we adopt the Sentence Transformer\footnote{\url{https://www.sbert.net/index.html}} models, i.e., 'all-MiniLM-L6-v2' for Event2012 and 'distiluse-base-multilingual-cased-v1' for Event2018. We use a 64 core Intel Xeon CPU E5-2680 with 512GB RAM and 1×NVIDIA Tesla P100-PICE GPU and report the mean over 5 runs for all experiments. 

\noindent\textbf{Evaluation Metrics.} 
%We measure adjusted mutual information (AMI) and adjusted rand index (ARI), which are broadly used by the previous studies \cite{cao2021knowledge}. 
We measure adjusted mutual information (AMI), adjusted rand index (ARI), and normalized mutual information (NMI, in Appendix), which are broadly used by the previous studies \cite{cao2021knowledge}. 
%Note that AMI is preferred over the commonly adopted Normalized Mutual Information (NMI), as the former is adjusted for chance, i.e., it accounts for the fact that the MI is generally higher for two clusterings with a larger number of clusters, regardless of whether there is actually more information shared.
%We report NMI in Appendix \ref{sec:nmis} (Tables \ref{table:closeset_nmi} - \ref{table:openset_nmi_event2018}), where similar observations as in \autoref{sec:overall_performance} - \ref{sec:hyper_parameter} hold and \framework~again outperforms all baselines.

\subsection{Overall Performance}\label{sec:overall_performance}
Tables 1 - 3 show the social event detection performance. 
\framework~consistently outperforms the highest baseline by large margins on both datasets across the closed- and open-set settings. 
%E.g., \framework~improves averaged ARI and AMI by 77\% and 19\% on Event2018 upon SBERT, the strongest baseline. 
E.g., on Event2018,~\framework~improves ARI and AMI upon SBERT, the strongest baseline, by 175\% and 6\% under the closed-set setting and by 77\% and 19\% on average under the open-set setting. This verifies that \framework~better explores the message semantics and the social network structure. 
Meanwhile, a comparison between the baselines indicates that the quality of the message embeddings matter: SBERT
%, fine-tuned for more accurate sentence similarity measurement \cite{reimers2019sentence}, 
outperforms BERT and the GNN-based methods.
Besides being effective and unsupervised, \framework~does not require predetermining the number of events. 
%This is essential as the exact number of events varies and is difficult to predict.
This is essential as the number of events is difficult to predict.
E.g., in Table 2, the ground truth number of events varies from 27 to 57 and can drop from 42 in $M_{15}$ to 27 in $M_{16}$ and raise from 30 in $M_{5}$ to 44 in $M_{6}$ between consecutive periods.
In contrast, KPGNN and QSGNN require labeled samples while KPGNN, QSGNN, BERT, and SBERT need the total number of events given a priori, which is impossible in practice. 
In short,~\framework~is more practical than the baselines and is the new SOTA.

\subsection{Ablation Study}
\label{sec:ablation_study}
Table 4 presents the ablation studies on Event2012. 
%$-\mathcal{E}_s$ removes the semantic-similarity-based $\mathcal{E}_s$ and simply relies on the common-attribute-based $\mathcal{E}_a$ to capture message correlations. Similarly, $-\mathcal{E}_a$ relies solely on $\mathcal{E}_s$ (unlike in \autoref{sec:1D_SE}, here we use the global rather than the first stable points since $\mathcal{E}_s$ is no longer a supplementation but aims to fully capture the message correlations). 
%\framework-BERT uses BERT rather than SBERT to measure the edge weights. \framework-vanilla partitions the message graph via vanilla 2D SE minimization (Algorithm \ref{algorithm:2D_SE}) instead of our proposed hierarchical one (Algorithm \ref{algorithm:hierarchical_2D_SE}). 
All the components of \framework~help. 
Especially, $\mathcal{E}_s$, absent in KPGNN and QSGNN, is essential for \framework's good performance. E.g., $-\mathcal{E}_s$ underperforms~\framework~by 52\% and 28\% in ARI and AMI, respectively, in the closed-set experiment. Also note \framework-BERT significantly outperforms BERT (shown in Tables 1 and 2), indicating that \framework~works despite the choice of PLM. Meanwhile, we observe that \framework-BERT underperforms \framework, indicating that PLM embeddings that are of High-quality and, in particular, faithfully reflect messages' semantic similarities (i.e., SBERT embeddings) are indispensable for \framework's good performance. Adopting embedding unsuitable for message similarity measuring (i.e., BERT embeddings), on the other hand, can lead to a decrease in performance (further discussed in Appendix).
A comparison to \framework-vanilla shows that \framework, adopts the proposed hierarchical 2D SE minimization algorithm, significantly improves efficiency without sacrificing performance: it performs on par with \framework-vanilla but is orders of magnitude faster (discussed in Section 3.3 and verified in Section 4.4).

\subsection{Efficiency of \framework}
\label{sec:efficiency}
\begin{figure}[t]
\centering
\includegraphics[width = 6.5cm]{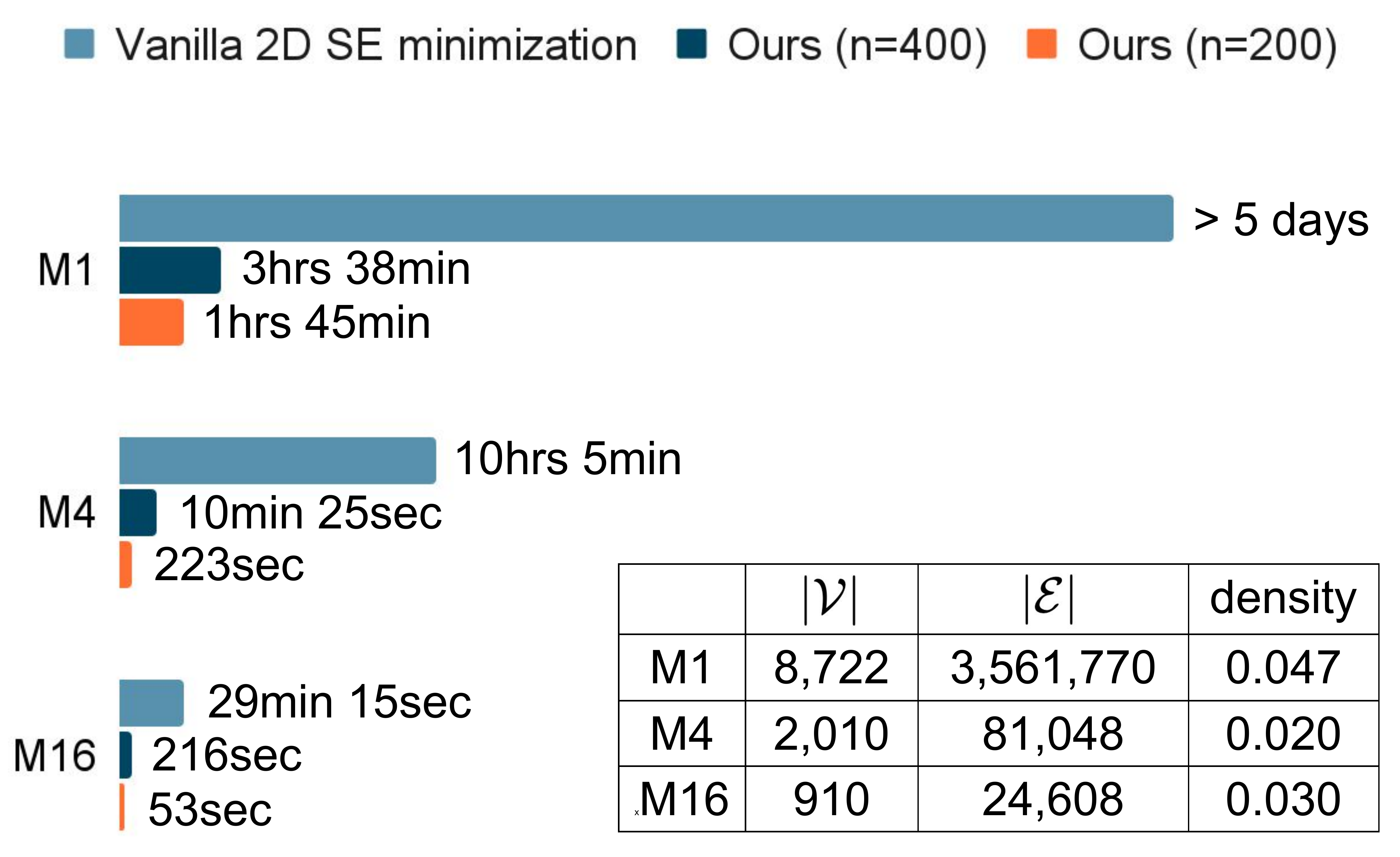}
\caption{Running time comparison between vanilla and hierarchical (ours) 2D SE minimization on Event2012.}
\label{fig:efficiency}
\end{figure}

We compare the efficiency of the proposed hierarchical 2D SE minimization to its vanilla predecessor. 
Figure 2 shows their time consumption on Event2012 message blocks. 
The vanilla algorithm runs prohibitively slow on complex message graphs. E.g., for a large and dense message block such as $M_1$, it takes more than 5 days to complete. In contrast, our proposed hierarchical 2D SE minimization dramatically reduces time consumption. E.g., for $M_1$, our algorithm reduces the running time by $>$97\%.
Adopting a smaller sub-graph size $n$ further decreases the running time. E.g., adopting a $n$ of 200 rather than 400 further reduces the time needed for $M_1$ by half. Also note that the impact to the performance when $n$ is decreased is rather small (Section 4.5).

\subsection{Hyperparameter Sensitivity}
\label{sec:hyper_parameter}

\begin{figure}[t]
\centering
\includegraphics[width = 6.3cm]{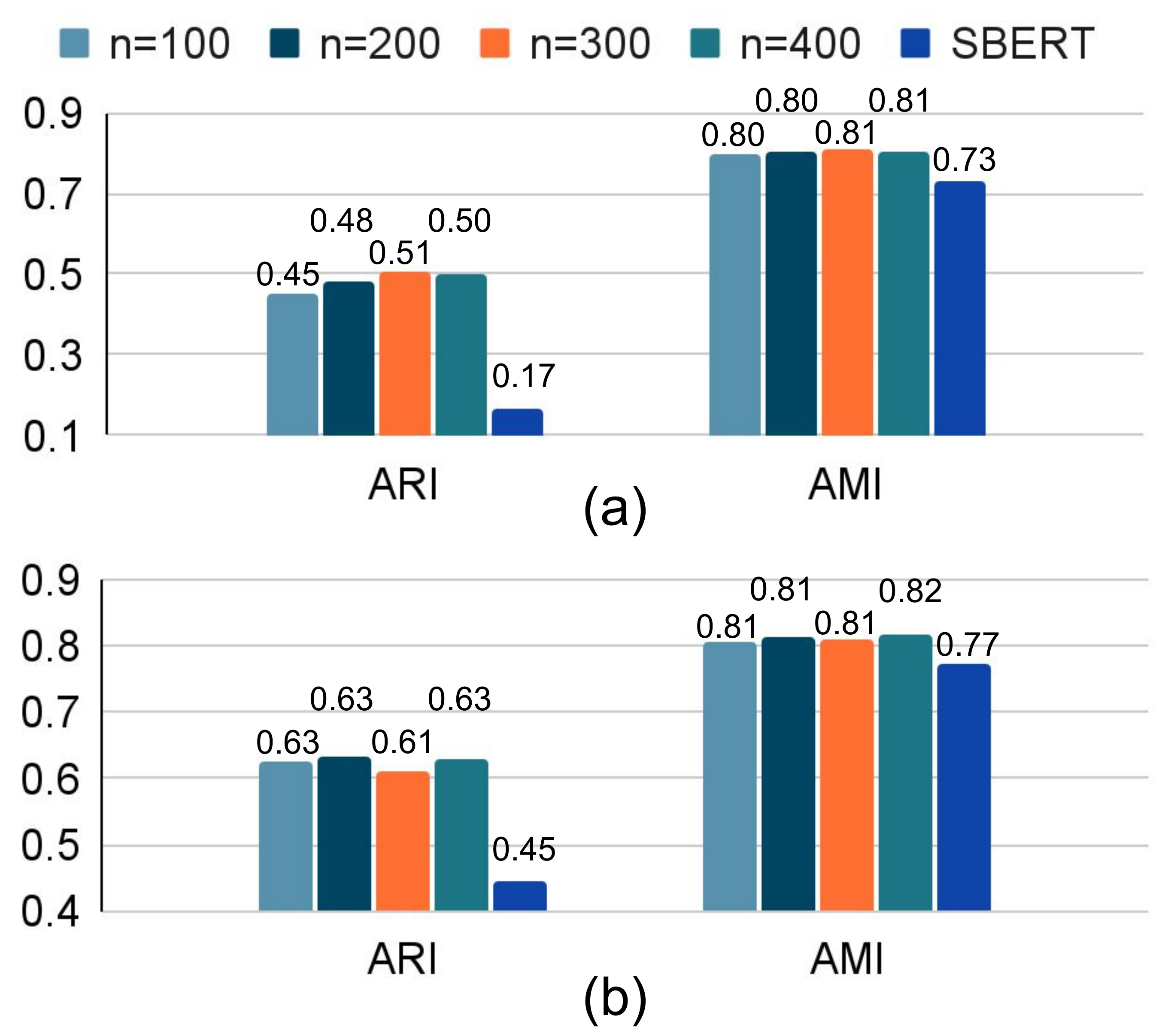}
\caption{\framework~results on Event2012 with different $n$. \textbf{(a)} and \textbf{(b)} show the closed-set and open-set (averaged) results.}
\label{fig:hyperparameter}
\end{figure}

We study how changing the sub-graph size $n$ affects the performance of \framework. Figure 3 shows that \framework~is relatively robust to the changes in $n$: increasing $n$ slightly prompts the performance at the cost of longer running time. 
Take Figure 3(a), the closed-set results on Event2012, for example, increasing $n$ from 100 to 400 introduces moderate (10\%) and marginal (1\%) improvements in ARI and AMI. 
Moreover, despite the changes in $n$, \framework~always outperforms SBERT, the strongest baseline, by 169-197\% in ARI and 9-10\% in AMI.

\subsection{Case Study}
\label{sec:case_study}
\begin{figure}[t]
\centering
\includegraphics[width = 7cm]{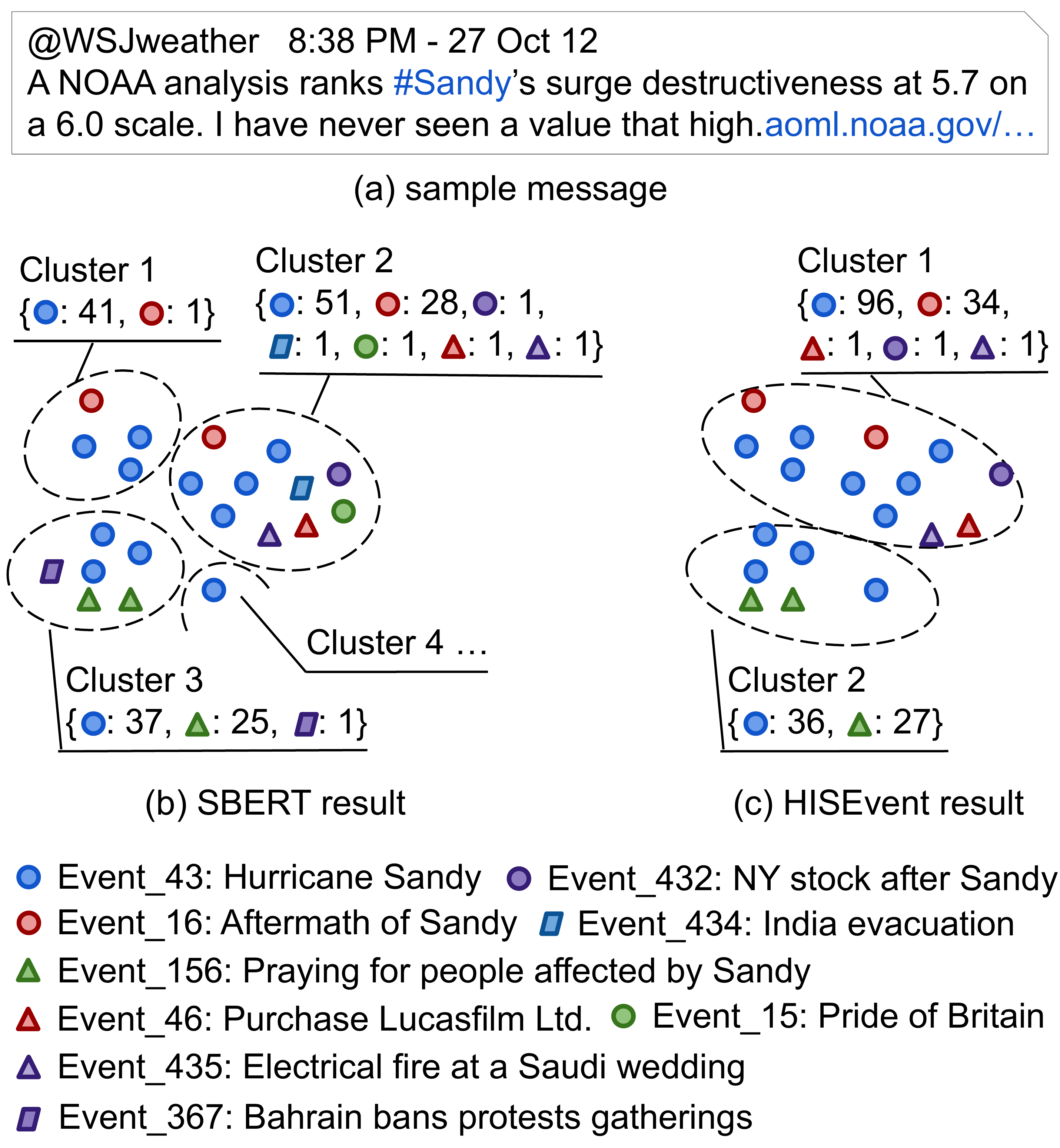}
\caption{Detection of \textit{Event\_43 Hurricane Sandy}. \textbf{(a)} is a sample message. \textbf{(b)} and \textbf{(c)} are clusters detected by SBERT and~\framework~that contain the target event messages.}
\label{fig:case_study}
\end{figure}

Figure 4 presents the detection of \textit{Event\_43 Hurricane Sandy}. We observe that the strongest baseline, SBERT, confuses the target event with many irrelevant events such as \textit{Event\_15 Pride of Britain} and \textit{Event\_367 Bahrain bans protests gathering}. As a result, the meaning of its detected clusters are rather vague. Moreover, SBERT outputs disjoint rather than more favorable, coherent clusters to represent the target event. E.g., it represents the target event with more than 4 clusters. In contrast, the proposed~\framework~detects justifiable clusters with clear meanings. E.g., cluster 1 is the destruction brought by Sandy while cluster 2 reflects the public reaction and recovery afterwards. Moreover,~\framework~fails only on the hard negatives. E.g., cluster 2 includes some messages about \textit{Event\_432 NY stock after Sandy}, which is relevant to the target event. 

\section{Related Work}
%goswami2016survey
 Social event detection is a long-standing task \cite{atefeh2015survey}. 
The main challenges lie in exploring the high-volume, complex, noisy, and dynamic social media components, e.g., text, timestamp, user mention, and social network structure. 
%hu2017adaptive, ozdikis2017incremental
Studies leverage incremental clustering \cite{zhao2007temporal, weng2011event, aggarwal2012event, zhang2007new, feng2015streamcube, xie2016topicsketch}, community detection \cite{fedoryszak2019real, liu2020story, liu2020event, yu2017ring}, and topic modeling \cite{zhou2014event, zhou2015unsupervised, xing2016hashtag, wang2016using, zhao2011comparing} are common. There are also methods for specific domains \cite{yao2020weakly, arachie2020unsupervised, khandpur2017determining} such as airport threats. 
They extract attributes, e.g., hashtag, from the social media components then pre-process the attributes in highly-customized manners.
%, often involving domain experts.
%cui2021mvgan
GNN-based methods \cite{peng2019fine, cao2021knowledge, peng2021streaming, ren2022evidential, ren2021transferring, ren2022known, peng2022reinforced} unify the various components concisely by introducing message graphs and quickly became a new trend for their outstanding performance. %\cite{peng2019fine} works in an offline (i.e., transductive) manner. \cite{cao2021knowledge} extends \cite{peng2019fine} to the more practical, online (i.e., inductive) setting. 
%The follow-up studies alleviate GNN's reliance on labeled samples by generating pseudo labels \cite{ren2022known} and seek to address cross-lingual scenarios \cite{peng2022reinforced}. 
\framework~keeps the merits of the GNN-based methods, better captures semantic-based message correlations, and eliminates sample labeling. 
Please also note that \textit{social event detection}, which highlights significant occurrences on social media, \textit{news story discovery} \cite{yoon2023unsupervised}, which summarizes long, formal, and plain textual news documents other than short, informal, and structural social messages, \textit{event prediction} \cite{zhao2018distant, deng2019learning, pan2020variational}, which forecasts future events, and \textit{event extraction} \cite{liu2019open}, which detects the entities, triggers, arguments, etc., of events, are non-comparable tasks.

%\noindent\textbf{Structural Entropy.} 
%Structural entropy (SE) \cite{li2016structural} measures  the amount of the structural information contained in a graph. Its minimization or maximization helps to disclose or obfuscate the underlying graph structure and is used in biomedical studies \cite{li2016three, li2018decoding}, privacy preservation \cite{liu2019rem}, graph pooling \cite{wu2022structural}, graph contrastive learning \cite{wu2023sega}, etc. \cite{liu2019rem} maximizes residual SE to select non-edges that best decept graph communities. \cite{li2016three} minimizes 1D SE to select reliable gene correlations. We also leverage 1D SE minimization to identify message correlations but in an efficient, incremental manner. \cite{wu2022structural} minimizes 2D SE and uses the resulting encoding trees, naturally hierarchical, as graph pooling output. They consider the original graph as a whole. In contrast, we customize our algorithm for the large-scale, dense social message graphs and perform highly efficient hierarchical minimization with repeated partitions and combinations. Concerning minimizing SE on dynamic graphs, \cite{yang22dynamic} updates the encoding tree upon the arrival of new edges and nodes. \framework, in contrast, is customized for social event detection and considers the hierarchical combination of clusters.

\section{Conclusion}
We address social event detection from a structural entropy perspective.~\framework~provides an effective, efficient, and unsupervised tool for social event detection and analysis. It keeps the merits of the GNN-based methods, better explores message correlations, and eliminates the need for labeling or predetermining the number of events. 
%\framework~addresses retrospective scenarios and can also be easily extended to streaming scenarios. We illustrate how~\framework~scales to real-world applications such as post-disaster response and public opinion monitoring, which can intrigue rescue workers, government personnel, and media practitioners. 
Experiments show that \framework~achieves the new SOTA under both closed- and open-set settings while being efficient and robust. 
%For future work, we plan to 1) apply~\framework~in streaming scenarios and explore its potentials in tracing the emergence of new events and the evolution of the existing events. 2) extend~\framework~to consider multi-relations by weighting the edges in the message graph differently, according to their types. 3) extend~\framework~to address higher-order (three or above) message graph SE minimization, which enables the study of inter-event correlations.

\section*{Acknowledgments}
The corresponding author is Hao Peng. 
This work is supported by National Key R\&D Program of China through grant 2022YFB3104700, NSFC through grants 62322202, U21B2027, 61972186, U23A20388 and 62266028, Beijing Natural Science Foundation through grant 4222030, Yunnan Provincial Major Science and Technology Special Plan Projects through grants 202302AD080003, 202202AD080003 and 202303AP140008, General Projects of  Basic Research in Yunnan Province through grants 202301AS070047 and 202301AT070471, and the Fundamental Research Funds for the Central Universities. 
Philip S. Yu was supported in part by NSF under grant III-2106758.

\bibliography{aaai24}

\appendix
\setcounter{secnumdepth}{2} %May be changed to 1 or 2 if section numbers are desired.

\section{Notations}
Table 5 summarizes the main notations used in this paper.

\begin{table}[t]
    \resizebox{\linewidth}{!}{%
    \begin{tabular}{r|l}  
    \hline
      \textbf{Notation} & \textbf{Description}\\
      \hline
      $G$ & Message graph   \\ 
      $\mathcal{V};\mathcal{E}$ & Node set of $G$; Edge set of $G$\\
      $\mathcal{T}$ & Encoding tree\\
      $\alpha, \lambda, \gamma \in \mathcal{T}$ & Node, root node, leaf node in $\mathcal{T}$\\
      $\alpha^-$ & The parent of $\alpha$\\
      $T_\alpha, T_\lambda, T_\gamma \in \mathcal{V}$ & Node sets $\subseteq \mathcal{V}$ that associate with  $\alpha, \lambda, \gamma$\\
      $h(\alpha);h(\mathcal{T})$ & Height of $\alpha$; Height of $\mathcal{T}$\\
      $g_\alpha$ & Summation of the degrees of the cut edges of $T_\alpha$\\
      $vol(\alpha);vol(\lambda)$ & Volume of $T_\alpha$; Volume of $T_\lambda$\\
      $\mathcal{H}^{(d)}$ & $d$-dimensional structural entropy (SE) of $G$\\
      $m$ & Message\\
      $e$ & Event\\
      $A_i$ & Attributes of message $m_i$\\
      $\boldsymbol{h}_{m_i}$ & PLM embedding of message $m_i$\\
      $\mathcal{E_{\text{a}}}$ & Common-attribute-based edges\\
      $\mathcal{E_{\text{s}}}$ & Semantic-similarity-based edges\\
      $d$ & Degree of a node in $G$\\
      $a_k$ & Nodes whose degrees altered by the $k$-NN edge set\\
      $n$ & Sub-graph size\\
      $\mathcal{P}$ & A partition of $\mathcal{V}$\\
      $\mathcal{P}_s$ & A subset of $\mathcal{P}$\\
      \hline
    \end{tabular}}
    \caption{Glossary of Notations.}
\end{table}

\section{Greedy Algorithm for 2D SE Minimization}
\label{sec:app_greedy_2D_SE_minimization}
\citealt{li2016structural} proposes a greedy algorithm for 2D SE minimization. Firstly, a MERGE operator is defined as follows:
\begin{definition}
\cite{li2016structural}. Given an encoding tree $\mathcal{T}$ and its two non-root nodes, $\alpha_{o_1}$ and $\alpha_{o_2}$, MERGE$(\alpha_{o_1}, \alpha_{o_2})$ removes $\alpha_{o_1}$ and $\alpha_{o_2}$ from $\mathcal{T}$ and adds a new node $\alpha_{n}$ to $\mathcal{T}$. $\alpha_{n}$ satisfies: 1) the children nodes of $\alpha_{n}$ in $\mathcal{T}$ is a combination of the children of $\alpha_{o_1}$ and $\alpha_{o_2}$; 2) ${\alpha_n}^- = \lambda$.
\end{definition}

The merge operation changes $\mathcal{T}$ and, therefore, would cause a change in 2D SE. 
Based on Definition 2, the change follows:
\begin{equation} \tag{4}
\label{eq:merge_delta_SE}
\begin{split}
& \Delta\text{SE}_{\alpha_{o_1}, \alpha_{o_2}} = \text{SE}_{new} -\text{SE}_{old} \\
& = - \frac{g_{\alpha_{n}}}{vol(\lambda)}log\frac{vol(\alpha_{n})}{vol(\lambda)} - \frac{vol(\alpha_{o_1})}{vol(\lambda)}log\frac{vol(\alpha_{o_1})}{vol(\alpha_{n})}\\
& - \frac{vol(\alpha_{o_2})}{vol(\lambda)}log\frac{vol(\alpha_{o_2})}{vol(\alpha_{n})} + \frac{g_{\alpha_{o_1}}}{vol(\lambda)}log\frac{vol(\alpha_{o_1})}{vol(\lambda)} \\
& + \frac{g_{\alpha_{o_2}}}{vol(\lambda)}log\frac{vol(\alpha_{o_2})}{vol(\lambda)}.
\end{split}
\end{equation}

The derivation of Equation 4 is in Appendix C. 
2D SE minimization can then be achieved by greedily and repeatedly merging the two nodes in $\mathcal{T}$ that would result in the largest $|\Delta \text{SE}|$ until no further merge can lead to a $\Delta \text{SE} < 0$. 
Algorithm 3 illustrates this process.

\begin{algorithm}
%\small
\caption{Greedy 2D SE minimization}\label{algorithm:2D_SE}
\KwIn{Message graph $G = (\mathcal{V}$, $\mathcal{E})$}
\KwOut{A partition $\mathcal{P}$ of $\mathcal{V}$}
Initialize $\mathcal{T}$, s.t. for each $m \in \mathcal{V}$, add two nodes, i.e., $\alpha$ that satisfies $T(\alpha) = \{m\}$ and ${\alpha}^-$, which is the parent of $\alpha$ and satisfies $h({\alpha}^-) = 1$, to $\mathcal{T}$\\
%Construct an initial $\mathcal{T}$, s.t. each $m_i \in \mathcal{V}$ is a leaf node $\alpha_i$ in $\mathcal{T}$ and $h(\alpha_i) = 1$\\
%$\mathcal{P} \gets (\alpha_i)^{\mathcal{V}}_{i=1}$\\
\While{True}{
    $\mathcal{P} \gets (\alpha|\alpha \in \mathcal{T}, h(\alpha) = 1)$\\
    $\Delta \text{SE} \gets \infty$\\
    \For{$\alpha_i \in \mathcal{P}$}{
        \For{$\alpha_j \in \mathcal{P}, j > i$}{
             $\Delta \text{SE}_{ij} \gets$ Eq. 4, w/o actually merging $\alpha_i$ and $\alpha_j$\\
            \If{$\Delta \text{SE}_{ij} < \Delta \text{SE}$}{
                $\Delta \text{SE} = \Delta \text{SE}_{ij}$\\
                $\alpha_{o_1} = \alpha_i$\\
                $\alpha_{o_2} = \alpha_j$
            }
        }
    }
    \If{$\Delta \text{SE} < 0$}{
        MERGE$(\alpha_{o_1}, \alpha_{o_2})$ 
    }\Else{
        Break
    }
}
\Return{$\mathcal{P}$}
\end{algorithm}

\section{Derivation of Equations 3 and 4}
\label{sec:app_derivation}

Equation 3 can be derived as follows:
\begin{equation} \tag{5}
\label{eq:increment_1D_de1}
\begin{split}
\mathcal{H}^{(1)\prime}(G) & = -\sum_{i=1}^{|\mathcal{V}|}\frac{d_i^\prime}{vol^\prime(\lambda)}log\frac{d_{i}^{\prime}}{vol^{\prime}(\lambda)} \\
& = -\sum_{i=1}^{|\mathcal{V}-a_{k}|}\frac{d_{i}}{vol^{\prime}(\lambda)}log\frac{d_{i}}{vol^{\prime}(\lambda)} \\
& -\sum_{j=1}^{|a_{k}|}\frac{d_{j}^{\prime}}{vol^{\prime}(\lambda)}log\frac{d_{j}^{\prime}}{vol^{\prime}(\lambda)}\\
& = \underbrace{-\sum_{i=1}^{|\mathcal{V}|}\frac{d_{i}}{vol^{\prime}(\lambda)}log\frac{d_{i}}{vol^{\prime}(\lambda)}}_{\tcircle{1}}\\
& + \sum_{j=1}^{|a_k|}\frac{d_j}{vol^\prime(\lambda)}log\frac{d_j}{vol^\prime(\lambda)}\\
& -\sum_{j=1}^{|a_k|}\frac{d_j^\prime}{vol^\prime(\lambda)}log\frac{d_j^\prime}{vol^\prime(\lambda)}.
\end{split}
\end{equation}

\begin{equation} \tag{6}
%\begin{multline}
\label{eq:increment_1D_de2}
\begin{split}
\tcircle{1} & = -\sum_{i=1}^{|\mathcal{V}|}\frac{d_i}{vol(\lambda)}\frac{vol(\lambda)}{vol^\prime(\lambda)}\Bigl(log\frac{d_i}{vol(\lambda)}+log\frac{vol(\lambda)}{vol^\prime(\lambda)}\Bigr) \\
& = - \frac{vol(\lambda)}{vol^\prime(\lambda)}\sum_{i=1}^{|\mathcal{V}|}\frac{d_i}{vol(\lambda)}log\frac{d_i}{vol(\lambda)} \\
& - \frac{vol(\lambda)}{vol^\prime(\lambda)}log\frac{vol(\lambda)}{vol^\prime(\lambda)}\sum_{i=1}^{|\mathcal{V}|}\frac{d_i}{vol(\lambda)}\\
& = \frac{vol(\lambda)}{vol^\prime(\lambda)}\Bigl(-\sum_{i=1}^{|\mathcal{V}|}\frac{d_i}{vol(\lambda)}log\frac{d_i}{vol(\lambda)} - log\frac{vol(\lambda)}{vol^\prime(\lambda)}\Bigr) \\
& = \frac{vol(\lambda)}{vol^\prime(\lambda)}\Bigl( \mathcal{H}^{(1)}(G) - log\frac{vol(\lambda)}{vol^\prime(\lambda)}\Bigr) .
\end{split}
\end{equation}

Plugging Equation 6 into Equation 5 concludes the derivation of Equation 3. 
In the above equations, $d_i$ denotes the original degree of a node $i$ in $G$ (initially, $d_i$ is calculated with $i$ linking to its 1st nearest neighbor).
$d_i^\prime$ denotes the updated degree of $i$ with an edge between $i$ and its $k$-th nearest neighbor inserted into $G$. 
$a_k$ is a set of nodes whose degree is affected by the insertion of the $k$-NN edge set. 
$vol(\lambda)$ and $vol^\prime(\lambda)$ stand for the volumes of $G$ before and after inserting the $k$-NN edge set. 
$\mathcal{H}^{(1)}(G)$ and $\mathcal{H}^{(1)\prime}(G)$ stand for the original and updated 1D SE. 

Equation 4 can be derived as follows:

\begin{equation} \tag{7}
\label{eq:merge_delta_SE_de1}
\begin{split}
& \Delta\text{SE}_{\alpha_{o_1}, \alpha_{o_2}} = \text{SE}_{new} -\text{SE}_{old} \\
& = - \frac{g_{\alpha_{n}}}{vol(\lambda)}log\frac{vol(\alpha_{n})}{vol(\lambda)} \underbrace{-\overset{|\Gamma_3|}{\underset{i=1}{\sum}}\frac{d_{\Gamma_{3i}}}{vol(\lambda)}log\frac{d_{\Gamma_{3i}}}{vol(\alpha_{n})}}_{\tcircle{1}}\\
& + \frac{g_{\alpha_{o_1}}}{vol(\lambda)}log\frac{vol(\alpha_{o_1})}{vol(\lambda)} \underbrace{+\overset{|\Gamma_{1}|}{\underset{i=1}{\sum}}\frac{d_{\Gamma_{1i}}}{vol(\lambda)}log\frac{d_{\Gamma_{1i}}}{vol(\alpha_{o_1})}}_{\tcircle{2}}\\
& + \frac{g_{\alpha_{o_2}}}{vol(\lambda)}log\frac{vol(\alpha_{o_2})}{vol(\lambda)} \underbrace{+\overset{|\Gamma_2|}{\underset{i=1}{\sum}}\frac{d_{\Gamma_{2i}}}{vol(\lambda)}log\frac{d_{\Gamma_{2i}}}{vol(\alpha_{o_2})}}_{\tcircle{3}},
\end{split}
\end{equation}

where $\Gamma_1 = \{\gamma|\gamma \in \mathcal{T}, \gamma^- = \alpha_{o_1}\}$, $\Gamma_2 = \{\gamma|\gamma \in \mathcal{T}, \gamma^- = \alpha_{o_2}\}$, and $\Gamma_3 = \{\gamma|\gamma \in \mathcal{T}, \gamma^- = \alpha_{n}\} = \Gamma_1 \cup \Gamma_2$ are sets of children nodes of $\alpha_{o_1}$, $\alpha_{o_2}$, and $\alpha_{n}$, respectively. Further, we have:

\begin{equation} \tag{8}
\label{eq:merge_delta_SE_de2}
\begin{split}
\tcircle{1} + \tcircle{2} + \tcircle{3} & = - \frac{vol(\alpha_{o_1})}{vol(\lambda)}log\frac{vol(\alpha_{o_1})}{vol(\alpha_{n})} \\
& - \frac{vol(\alpha_{o_2})}{vol(\lambda)}log\frac{vol(\alpha_{o_2})}{vol(\alpha_{n})}.
\end{split}
\end{equation}

Plugging Equation 7 into Equation 8 concludes the derivation of Equation 4.

\section{Dataset Splits}
\label{sec:data_splits}
Although the proposed \framework~is fully unsupervised, some of the baseline methods require training. 
We, therefore, follow the data splits as adopted by those baselines \cite{cao2021knowledge, ren2022known}. 
Tables 6, 7, and 8 show the statistics of the data splits.
For the close-set (offline) scenario, which simultaneously considers all the event classes, the two datasets are randomly split by 70:10:20 into training, validation, and test sets. 
On the other hand, the open-set (online) scenario considers how events happen over time and splits the datasets into temporal message blocks. 
The messages of the first week form an initial block ($M_0$) for training and validation, while the successive day-wise message blocks ($M_1$ through $M_{21}$ for Event2012 and $M_1$ through $M_{16}$ for Event2018) are used for testing. 

\begin{table*}[t]
\small
%\scriptsize
\centering
    \begin{tabular}{c|ccc|ccc}
    \hline
    \multirow{2}{*}{Dataset} & \multicolumn{3}{c|}{Event2012} & \multicolumn{3}{c}{Event2018}\\
    %\cline{2-7}
     & Train & Val & Test & Train & Val & Test \\
     \hline
     $\#$ messages & 48,188 & 6,884 & 13,769 & 45,162 & 6,452 & 12,902 \\
     $\#$ events & 499 & 461 & 488 & 255 & 225 & 241 \\
    \hline
    \end{tabular}
    \caption{Close-set data splits.}
  \label{table:Closeset_data_splits}
\end{table*}

\begin{table*}[t]
\small
  \centering
    \begin{tabular}{c|ccccccccccc}
   \hline
    Blocks & $M_0$ & $M_1$ & $M_2$ & $M_3$ & $M_4$ & $M_5$ & $M_6$ & $M_7$ & $M_8$ & $M_9$ & $M_{10}$\\
    \hline
    $\#$ messages & 20,254 & 8,722 & 1,491 & 1,835 & 2,010 & 1,834 & 1,276 & 5,278 & 1,560 & 1,363 & 1,096 \\ 
    $\#$ events & 155 & 41 & 30 & 33 & 38 & 30 & 44 & 57 & 53 & 38 & 33 \\ 
    \hline
    \hline
    Blocks & $M_{11}$ & $M_{12}$ & $M_{13}$ & $M_{14}$ & $M_{15}$ & $M_{16}$ & $M_{17}$ & $M_{18}$ & $M_{19}$ & $M_{20}$ & $M_{21}$ \\ 
    \hline
    $\#$ messages & 1,232 & 3,237 & 1,972 & 2,956 & 2,549 & 910 & 2,676 & 1,887 & 1,399 & 893 & 2,410 \\ 
    $\#$ events & 30 & 42 & 40 & 43 & 42 & 27 & 35 & 32 & 28 & 34 & 32 \\ 
    \hline
  \end{tabular}
  \caption{Open-set data splits of Event2012.}
  \label{table:Openset_data_splits_Event2012}
\end{table*}

\begin{table*}[t]
\small
  \centering
    \begin{tabular}{c|ccccccccc}
   \hline
    Blocks & $M_0$ & $M_1$ & $M_2$ & $M_3$ & $M_4$ & $M_5$ & $M_6$ & $M_7$ & $M_8$ \\
    \hline
    $\#$ messages & 14,328 & 5,356 & 3,186 & 2,644 & 3,179 & 2,662 & 4,200 & 3,454 & 2,257 \\ 
    $\#$ events & 79 & 22 & 19 & 15 & 19 & 27 & 26 & 23 & 25\\ 
    \hline
    \hline
    Blocks & $M_9$ & $M_{10}$ & $M_{11}$ & $M_{12}$ & $M_{13}$ & $M_{14}$ & $M_{15}$ & $M_{16}$ & \\
    \hline
    $\#$ messages & 3,669 & 2,385 & 2,802 & 2,927 & 4,884 & 3,065 & 2,411 & 1,107 & \\ 
    $\#$ events & 31 & 32 & 31 & 29 & 28 & 26 & 25 & 14 & \\ 
    \hline
  \end{tabular}
  \caption{Open-set data splits of Event2018.}
  \label{table:Openset_data_splits_Event2018}
\end{table*}

\section{Experiment Setting} 
For \framework, we adopt SBERT to calculate edge weights (the calculation follows Section 3.2 and the effects of changing the PLM are observed in Section 4.3). We set the sub-graph size $n$ for the closed-set experiments to 300 and 800 for Event2012 and Event2018, respectively. 
As to the open-set experiments, we set $n$ to 400 and 300 for Event2012 and Event2018 (the effects of changing $n$ are observed in Section 4.5). 
For KPGNN, QSGNN, and EventX, we adopt the settings as reported in the original papers. 
For BERT, we adopt Hugging Face\footnote{\url{https://huggingface.co/models?filter=bert}} pretrained models, i.e., 'bert-large-cased' for Event2012 and 'bert-base-multilingual-cased' for Event2018. 
We apply mean pooling (i.e., average the last hidden states of the words in a message for its embedding) as we empirically found it outperforms the pooler and [CLS] output. 
For SBERT, we adopt the Sentence Transformer\footnote{\url{https://www.sbert.net/index.html}} models, i.e., 'all-MiniLM-L6-v2' for Event2012 and 'distiluse-base-multilingual-cased-v1' for Event2018. We use a 64 core Intel Xeon CPU E5-2680 with 512GB RAM and 1×NVIDIA Tesla P100-PICE GPU and report the mean over 5 runs for all experiments.

\section{Social Event Detection NMIs}
\label{sec:nmis}
We report the NMI scores of the close- and open-set experiments on both datasets in Tables 9, 10, and 11. 
We can tell that the observations made in Section 4.2 hold. 

\begin{table*}[t]
\small
\centering
    \begin{tabular}{c|ccccc|cc}
    \hline
    Dataset & KPGNN* & QSGNN* & EventX & BERT* & SBERT* & \framework & Improv. (\%)\\
    \hline
    Event2012 & 0.70 & 0.72 & 0.72 & 0.55 & \textit{0.83} & \textbf{0.85} & $\uparrow$2 \\
    \hline
    Event2018 & 0.56 & 0.58 & 0.56 & 0.46 & \textit{0.69} & \textbf{0.70} & $\uparrow$1 \\
    \hline
    \end{tabular}
    \caption{Close-set NMIs. * marks results acquired with the ground truth event numbers.}
  \label{table:closeset_nmi}
\end{table*}

\begin{table*}[t]
\small
  \centering
    \begin{tabular}{c|ccccccccccc}
    \hline
    Blocks & $M_1$ & $M_2$ & $M_3$ & $M_4$ & $M_5$ & $M_6$ & $M_7$ & $M_8$ & $M_9$ & $M_{10}$ & $M_{11}$ \\
    \hline
    KPGNN* & 0.39 & 0.79 & 0.76 & 0.67 & 0.73 & 0.82 & 0.55 & 0.80 & 0.74 & 0.80 & 0.74\\ 
    QSGNN* & \textit{0.43} & 0.81 & 0.78 & 0.71 & \textit{0.75} & 0.83 & 0.57 & 0.79 & 0.77 & 0.82 & \textit{0.75}\\
    EventX & 0.36 & 0.68 & 0.63 & 0.63 & 0.59 & 0.70 & 0.51 & 0.71 & 0.67 & 0.68 & 0.65 \\
    BERT* & 0.37 & 0.78 & 0.75 & 0.62 & 0.70 & 0.79 & 0.53 & 0.78 & 0.75 & 0.80 & 0.66 \\
    SBERT* & 0.40 & \textit{0.86} & \textit{0.88} & \textit{0.82} & \textbf{0.86} & \textit{0.86} & \textit{0.64} & \textit{0.88} & \textit{0.85} & \textit{0.87} & \textbf{0.84} \\
    \hline
    \framework & \textbf{0.45} & \textbf{0.89} & \textbf{0.94} & \textbf{0.85} & \textbf{0.86} & \textbf{0.91} & \textbf{0.70} & \textbf{0.90} & \textbf{0.89} & \textbf{0.91} & \textbf{0.84} \\
    Improv. (\%) & $\uparrow$13 & $\uparrow$3 & $\uparrow$7 & $\uparrow$4 &$\rightarrow$ &$\uparrow$6 &$\uparrow$9 &$\uparrow$2 &$\uparrow$5 &$\uparrow$5 & $\rightarrow$ \\
    \hline
    \hline
    Blocks & $M_{12}$ & $M_{13}$ & $M_{14}$ & $M_{15}$ & $M_{16}$ & $M_{17}$ & $M_{18}$ & $M_{19}$ & $M_{20}$ & $M_{21}$ & Avg.\\ 
    \hline
    KPGNN* & 0.68 & 0.69 & 0.69 & 0.58 & 0.79 & 0.70 & 0.68 & 0.73 & 0.72 & 0.60 & 0.70 \\
    QSGNN* & 0.70 & 0.68 & 0.68 & 0.59 & 0.78 & 0.71 & 0.70 & 0.73 & \textit{0.73} & 0.61 & 0.71 \\
    EventX & 0.61 & 0.58 & 0.57 & 0.49 & 0.62 & 0.58 & 0.59 & 0.60 & 0.67 & 0.53 & 0.60 \\
    BERT* & 0.59 & 0.61 & 0.59 & 0.48 & 0.74 & 0.59 & 0.55 & 0.63 & 0.67 & 0.57 & 0.65 \\
    SBERT* & \textit{0.86} & \textit{0.73} & \textit{0.79} & \textit{0.70} & \textit{0.81} & \textit{0.78} & \textbf{0.82} & \textit{0.84} & \textbf{0.83} & \textbf{0.72} & \textit{0.79}\\
    \hline
    \framework & \textbf{0.90} & \textbf{0.79} & \textbf{0.88} & \textbf{0.74} & \textbf{0.88} & \textbf{0.82} & \textit{0.80} & \textbf{0.88} & \textbf{0.83} & \textit{0.70} & \textbf{0.83} \\
    Improv. (\%) & $\uparrow$5 & $\uparrow$8 & $\uparrow$11 & $\uparrow$6 & $\uparrow$9 & $\uparrow$5 & $\downarrow$2 & $\uparrow$5 & $\rightarrow$ & $\downarrow$3 & $\uparrow$5 \\
    \hline
  \end{tabular}
  \caption{Open-set NMIs on Event2012. * marks results acquired with the ground truth event numbers.}
  \label{table:openset_nmi_event2012}
\end{table*}

\begin{table*}[t]
\small
  \centering
    \begin{tabular}{c|ccccccccc}
    \hline
    Blocks & $M_1$ & $M_2$ & $M_3$ & $M_4$ & $M_5$ & $M_6$ & $M_7$ & $M_8$ & $M_9$ \\
    \hline
    KPGNN* & 0.54 & 0.56 & 0.52 & 0.55 & 0.58 & 0.59 & 0.63 & 0.58 & 0.48 \\ 
    QSGNN* & 0.57 & 0.58 & 0.57 & 0.58 & 0.61 & 0.60 & 0.64 & 0.57 & 0.52 \\
    EventX & 0.34 & 0.37 & 0.37 & 0.39 & 0.53 & 0.44 & 0.41 & 0.54 & 0.45 \\
    BERT* & 0.43 & 0.45 & 0.45 & 0.42 & 0.58 & 0.50 & 0.50 & 0.52 & 0.44 \\
    SBERT* & \textit{0.60} & \textit{0.62} & \textit{0.64} & \textit{0.61} & \textit{0.77} & \textit{0.73} & \textit{0.66} & \textit{0.76} & \textit{0.64} \\
    \hline
    \framework & \textbf{0.77} & \textbf{0.80} & \textbf{0.75} & \textbf{0.73} & \textbf{0.83} & \textbf{0.83} & \textbf{0.82} & \textbf{0.90} & \textbf{0.74} \\
    Improv. (\%) & $\uparrow$28 & $\uparrow$29 & $\uparrow$17 & $\uparrow$20 & $\uparrow$8 & $\uparrow$14 & $\uparrow$24 & $\uparrow$18 & $\uparrow$16 \\
    \hline
    \hline
    Blocks  & $M_{10}$ & $M_{11}$ & $M_{12}$ & $M_{13}$ & $M_{14}$ & $M_{15}$ & $M_{16}$ & Avg. & \\ 
    \hline
    KPGNN* & 0.57 & 0.54 & 0.55 & 0.60 & 0.66 & 0.60 & 0.52 & 0.57 \\
    QSGNN* & 0.60 & 0.60 & 0.61 & 0.59 & 0.68 & 0.63 & 0.51 & 0.59 \\
    EventX & 0.52 & 0.48 & 0.51 & 0.44 & 0.52 & 0.49 & 0.39 & 0.45 \\
    BERT* & 0.49 & 0.51 & 0.56 & 0.41 & 0.53 & 0.55 & 0.45 & 0.49 \\
    SBERT* & \textit{0.74} & \textit{0.72} & \textit{0.77} & \textit{0.66} & \textit{0.69} & \textit{0.72} & \textit{0.66} & \textit{0.69} \\
    \hline
    \framework & \textbf{0.81} & \textbf{0.80} & \textbf{0.89} & \textbf{0.89} & \textbf{0.90} & \textbf{0.84} & \textbf{0.74} & \textbf{0.81} \\
    Improv. (\%) & $\uparrow$9 & $\uparrow$11 & $\uparrow$16 & $\uparrow$35 & $\uparrow$30 & $\uparrow$17 & $\uparrow$12 & $\uparrow$17 \\
    \hline
  \end{tabular}
  \caption{Open-set NMIs on Event2018. * marks results acquired with the ground truth event numbers.}
  \label{table:openset_nmi_event2018}
\end{table*}

\section{Scale~\framework~to very large data}
\framework~scales to large-scale data as it can be easily parallelized. Specifically,~\framework~splits the original message graph into sub-graphs of size $n$, which can then be handled concurrently. Moreover, in real-world applications (e.g., post-disaster rescue), social streams are typically split (e.g., by locations) and filtered (e.g., to eliminate redundancy) before further processing, while the message graphs are kept small (e.g., only contain messages posted in the last hour) to ensure timely responses.~\framework~can thus be used in real-world monitoring applications such as emergency response after disasters and public opinion monitoring, which can intrigue rescue workers, government personnel, and media practitioners.

\section{Extend~\framework~to streaming scenarios}
\label{sec:extend}
\framework~addresses retrospective scenarios as it performs batched detection. I.e., it processes a batch of messages at a time and separates the events through clustering the messages in the batch. Hence,~\framework~is different from streaming social event detection methods \cite{aggarwal2012event} that detect emerging events through monitoring indicators such as the changes in the volume of some keywords. However, we would like to point out that~\framework~can be easily extended to streaming scenarios. Specifically, by choosing smaller batch sizes (e.g., constructing message graphs for each past hour or even minute), one can extract the latest events (as message clusters). To trace the emergence of new events and the evolution of the existing events, one can ‘chain up’ the message clusters from the consecutive batches, either according to their semantic similarities or by applying~\framework~to merge these clusters. 

\section{Limitations}
PLM embeddings that are of High-quality and, in particular, faithfully reflect messages' semantic similarities are indispensable for \framework's good performance. 
Adopting embedding unsuitable for message similarity measuring can lead to a decrease in the \framework's performance. 
For example, in Table 4, \framework-BERT scores lower than \framework. 
We recommend SBERT \cite{reimers2019sentence}, which is fine-tuned for sentence similarity measuring, to be used for the \framework's semantic-similarity-based edge ($\mathcal{E}_s$) selection and edge weight calculation. 
In multi-lingual and/or cross-lingual scenarios, the 'distills-base-multilingual-cased-v1' version of SBERT, which supports 15 languages, can be applied and has shown good performance in our experiments (Section 4.2). 
For languages that are unsupported by SBERT, alternative PLMs, e.g., XLM-RoBERTa \footnote{\url{https://huggingface.co/docs/transformers/model_doc/xlm-roberta}}, mT5 \footnote{\url{https://huggingface.co/docs/transformers/model_doc/mt5}}, etc., need to be decided, and \framework's performance may deteriorate.

\end{document}